\documentclass{emulateapj}

\long\def\symbolfootnote[#1]#2{\begingroup%
\def\thefootnote{\fnsymbol{footnote}}\footnote[#1]{#2}\endgroup}

\def\HII{\hbox{H\,{\sc ii}}}

\def\OIV{\hbox{[O\,{\sc iv}]25.89$\,\mu$m}}

\def\NeII{\hbox{[Ne\,{\sc ii}]12.81$\,\mu$m}}

\def\NeV{\hbox{[Ne\,{\sc v}]14.32$\,\mu$m}}

\def\CIno{\hbox{[C\,{\sc i}]}}
\def\CII{\hbox{[C\,{\sc ii}]157.7$\,\mu$m}}
\def\CIIno{\hbox{[C\,{\sc ii}]}}

\def\LCIIno{\hbox{$L_{\rm [C\,{\sc II}]}$}}

\def\OI{\hbox{[O\,{\sc i}]63.18$\,\mu$m}}
\def\OIno{\hbox{[O\,{\sc i}]}}
\def\OIb{\hbox{[O\,{\sc i}]145$\,\mu$m}}

\def\NIIa{\hbox{[N\,{\sc ii}]122$\,\mu$m}}

\def\OH{\hbox{OH\,79.18$\,\mu$m}}

\def\SSi{\hbox{$S_{\rm 9.7\,\mu m}$}}
\def\H2{\hbox{H$_{2}$}}

\def\PAHc{\hbox{6.2$\,\mu$m\,PAH}}

\def\deg{$^{\circ}$}

\def\kms{${\rm km~s}^{-1}$}

\def\nh{$n_{\rm H}$}
\def\Go{$G_{\rm 0}$}
\def\MHH{\hbox{$M_{\rm H_2}$}}
\def\Mgas{\hbox{$M_{\rm gas}$}}
\def\Lsun{\hbox{$L_\odot$}}
\def\Msun{\hbox{$M_{\odot}$}}
\def\Mstar{\hbox{$M_{\star}$}}
\def\LIR{\hbox{$L_{\rm IR}$}}
\def\LFIR{\hbox{$L_{\rm FIR}$}}

\def\c-m{\hbox{cm$^{-1}$}}
\def\c-mm{\hbox{cm$^{-2}$}}
\def\c-mmm{\hbox{cm$^{-3}$}}
\def\cmmm{\hbox{cm$^{3}$}}
\def\kms{\hbox{km$\,$s$^{-1}$}}

\def\wm{\hbox{W$\,$m$^{-2}$}}

\def\Tdust{\hbox{$T_{\rm dust}$}}

\shorttitle{Explaining the $\CIIno\,$ Deficit in LIRGs -- First GOALS Results From Herschel/PACS}

\shortauthors{D\'iaz-Santos et al.}

\begin{document}

\title{Explaining the $\CII\,$ Deficit in Luminous Infrared Galaxies -- First Results from a Herschel/PACS Study of the GOALS Sample}


\author{T.~D\'{\i}az-Santos\altaffilmark{1,\dag},
L.~Armus\altaffilmark{1},
V.~Charmandaris\altaffilmark{2,3},
S.~Stierwalt\altaffilmark{4},
E.~J.~Murphy\altaffilmark{5},
S.~Haan\altaffilmark{6},
H.~Inami\altaffilmark{7}
S.~Malhotra\altaffilmark{8},
R.~Meijerink\altaffilmark{9},
G.~Stacey\altaffilmark{10},
A.~O.~Petric\altaffilmark{11},
A.~S.~Evans\altaffilmark{4,12},
S.~Veilleux\altaffilmark{13,14},
P.~P.~van~der~Werf\altaffilmark{15},
S.~Lord\altaffilmark{16},
N.~Lu\altaffilmark{12,17},
J.~H.~Howell\altaffilmark{1},
P.~Appleton\altaffilmark{16},
J.~M.~Mazzarella\altaffilmark{4},
J.~A.~Surace\altaffilmark{1},
C.~K.~Xu\altaffilmark{16}
B.~Schulz\altaffilmark{12,17}
D.~B.~Sanders\altaffilmark{18},
C.~Bridge\altaffilmark{12},
B.~H.~P.~Chan\altaffilmark{12},
D.~T.~Frayer\altaffilmark{19},
K.~Iwasawa\altaffilmark{20},
J.~Melbourne\altaffilmark{21},
and E.~Sturm\altaffilmark{22}
}

\altaffiltext{\dag}{Contact email: tanio@ipac.caltech.edu}
\affil{$^{1}$Spitzer Science Center, California Institute of Technology, MS 220-6, Pasadena, 91125, CA}
\affil{$^{2}$IESL/Foundation for Research and Technology - Hellas, GR-71110, Heraklion, Greece and Chercheur Associ\'e, Observatoire de  Paris, F-75014, Paris, France}
\affil{$^{3}$University of Crete, Department of Physics, GR-71003, Heraklion}
\affil{$^{4}$Department of Astronomy, University of Virginia, P.O. Box 400325, Charlottesville, VA 22904}
\affil{$^{5}$Observatories of the Carnegie Institution for Science, 813 Santa Barbara Street, Pasadena, CA 91101, USA}
\affil{$^{6}$CSIRO Astronomy and Space Science, Marsfield NSW, 2122, Australia}
\affil{$^{7}$National Optical Astronomy Observatory, 950 N. Cherry Ave., Tucson, AZ 85719, USA}
\affil{$^{8}$School of Earth and Space Exploration, Arizona State University, Tempe, AZ 85287, USA}
\affil{$^{9}$Kapteyn Astronomical Institute, University of Groningen, P.O. Box 800, NL-9700 AV Groningen, The Netherlands}
\affil{$^{10}$Department of Astronomy, Cornell University, Ithaca, NY 14853, USA}
\affil{$^{11}$Astronomy Department, California Institute of Technology, Pasadena, CA 91125, USA}
\affil{$^{12}$National Radio Astronomy Observatory, 520 Edgemont Road, Charlottesville, VA 22903}
\affil{$^{13}$Joint Space-Science Institute, University of Maryland, College Park, MD 20742, USA}
\affil{$^{14}$Department of Astronomy, University of Maryland, College Park, MD 20742, USA}
\affil{$^{15}$Leiden Observatory, Leiden University, P.O. Box 9513, NL-2300 RA Leiden, The Netherlands}
\affil{$^{16}$NASA Herschel Science Center, IPAC, California Institute of Technology, MS 100-22, Cech, Pasadena, CA 91125}
\affil{$^{17}$Infrared Processing and Analysis Center, MS 100-22, California Institute of Technology, Pasadena, CA 91125}
\affil{$^{18}$Institute for Astronomy, University of Hawaii, 2680 Woodlawn Drive, Honolulu, HI 96822}
\affil{$^{19}$National Radio Astronomy Observatory, P.O. Box 2, Green Bank, WV 24944, USA}
\affil{$^{20}$ICREA and Institut de Cincies del Cosmos (ICC), Universitat de Barcelona (IEEC-UB), Marti i Franques 1, 08028 Barcelona, Spain}
\affil{$^{21}$Caltech Optical Observatories, Division of Physics, Mathematics and Astronomy, MS 301-17, California Institute of Technology, Pasadena, CA 91125, USA}
\affil{$^{22}$Max-Planck-Institut f\"{u}r extraterrestrische Physik, Postfach 1312, D-85741 Garching, Germany}

\begin{abstract}

We present the first results of a survey of the \CII\, emission line in 241 luminous infrared galaxies (LIRGs) comprising the Great Observatories All-sky Survey (GOALS) sample, obtained with the PACS instrument on board the \textit{Herschel Space Observatory}. The \CIIno\, luminosities, \LCIIno, of the LIRGs in GOALS range from $\sim\,10^7$ to 2$\,\times\,10^9\,$\Lsun. We find that LIRGs show a tight correlation of \CIIno/FIR\, with far-IR flux density ratios, with a strong negative trend spanning from $\sim\,10^{-2}$ to $10^{-4}$, as the average temperature of dust increases. We find correlations between the \CIIno/FIR\, ratio and the strength of the 9.7$\,\mu$m silicate absorption feature as well as with the luminosity surface density of the mid-IR emitting region ($\Sigma_{\rm MIR}$), suggesting that warmer, more compact starbursts have substantially smaller \CIIno/FIR\, ratios.
Pure star-forming LIRGs have a mean \CIIno/FIR\,$\sim\,4\,\times\,10^{-3}$, while galaxies with low \PAHc\, equivalent widths (EWs), indicative of the presence of active galactic nuclei (AGN), span the full range in \CIIno/FIR. However, we show that even when only pure star-forming galaxies are considered, the \CIIno/FIR\, ratio still drops by an order of magnitude, from $10^{-2}$ to $10^{-3}$, with $\Sigma_{\rm MIR}$ and $\Sigma_{\rm IR}$, implying that the \CII\, luminosity is not a good indicator of the star formation rate (SFR) for most LIRGs, for it does not scale linearly with the warm dust emission most likely associated to the youngest stars. Moreover, even in LIRGs in which we detect an AGN in the mid-IR, the majority (2/3) of galaxies show \CIIno/FIR\,$\geq\,10^{-3}$ typical of high \PAHc\, EW sources, suggesting that most AGNs do not contribute significantly to the far-IR emission.
We provide an empirical relation between the \CIIno/FIR\, and the specific SFR (SSFR) for star-forming LIRGs.
Finally, we present predictions for the starburst size based on the observed \CIIno\, and far-IR luminosities which should be useful for comparing with results from future surveys of  high-redshift galaxies with ALMA and CCAT.

\end{abstract}


\section{Introduction}\label{s:intro}


Systematic spectroscopic observations of far-infrared (IR) cooling lines in large samples of local star-forming galaxies and active galactic nuclei (AGN) were first carried out with \textit{ISO} (e.g., \citealt{Malhotra1997}, 2001; \citealt{Luhman1998}; \citealt{Brauher2008}). These studies showed that \CII\, is the most intense far-IR emission line observed in normal, star-forming galaxies (\citealt{Malhotra1997}) and starbursts (e.g., \citealt{Nikola1998}; \citealt{Colbert1999}), dominating the gas cooling of their neutral inter stellar medium (ISM). This fine-structure line arises from the $^2P_{3/2}\,\rightarrow\,^2P_{1/2}$ transition ($E_{ul}/k\,=\,92$\,K) of singley ionized Carbon atoms (ionization potential\,=\,11.26\,eV and critical density, $n^{\rm cr}_{\rm H}\,\simeq\,2.7\times 10^3\,\c-mmm$;  $n^{\rm cr}_{\rm e^-}\,\simeq\,46\,\c-mmm$) which are predominantly excited by collisions with neutral hydrogen atoms; or with free electrons and protons in regions where $n_{\rm e^-}/n_{\rm H}\,\gtrsim\,10^{-3}$ (\citealt{Hayes1984}). Ultraviolet (UV) photons with energies $>\,$6\,eV emitted by newly formed stars are able to release the most weakly bound electrons from small dust grains via photo-electric heating (\citealt{Watson1972}; \citealt{Draine1978}). In particular, polycyclic aromatic hydrocarbons (PAHs) are thought to be an important source of photo-electrons (\citealt{Helou2001}) that contribute, through kinetic energy transfer, to the heating of the neutral gas which subsequently cools down via collision with C$^+$ atoms and other elements in photo-dissociation regions (PDRs) (\citealt{Tielens1985}; \citealt{Wolfire1995}).


The \CII\, emission accounts, in the most extreme cases, for as much as $\sim\,1\,$\% of the total IR luminosity of galaxies (\citealt{Stacey1991}; \citealt{Helou2001}). However, the \CIIno/FIR\, ratio is observed to decrease by more than an order of magnitude in sources with high \LIR\, and warm dust temperatures (\Tdust). The underlying causes for these trends are still debated. The physical arguments most often proposed to explain the decrease in \CIIno/FIR\, are: (1) self-absorption of the C$^+$ emission, (2) saturation of the \CIIno\, line flux due to high density of the neutral gas, (3) progressive ionization of dust grains in high far-UV field to gas density environments, and (4) high dust-to-gas opacity caused by an increase of the average ionization parameter.

Although self absorption has been used to explain the faint \CIIno\, emission arising from warm, AGN-dominated systems such as Mrk~231 (\citealt{Fischer2010}), this interpretation has been questioned in normal star-forming galaxies due to the requirement of extraordinarly large column densities of gas in the PDRs (\citealt{Luhman1998}, \citealt{Malhotra2001}). Furthermore, contrary to the \OIno\, or \CIno\, lines, the \CIIno\, emission is observed to arise from the external edges of those molecular clouds exposed to the UV radiation originated from starbursts, as for example in Arp~220 (\citealt{Conti2012}). Therefore, self absorption is not the likely explanation of the low \CIIno/FIR\, ratios seen in most starburst galaxies, except perhaps in a few extreme cases, like NGC~4418 (\citealt{Malhotra1997}).
 
The \CIIno\, emission becomes saturated when the hydrogen density in the neutral medium, \nh, increases to values $\gtrsim\,10^3\,\c-mmm$, provided that the far-UV ($6-13.6\,$eV) radiation field is not extreme (\Go$\,\lesssim 10^4$; where \Go\, is normalized to the average local interstellar radiation field; \citealt{Habing1968}). For example, for a constant \Go$\,=\,10^2$, an increase of the gas density from $10^4$ to $10^6\,\c-mmm$ would produce a suppression of the \CIIno\, emission of almost 2 orders of magnitude due to the rapid recombination of C$^+$ into neutral Carbon and then into CO (\citealt{Kaufman1999}). However, PDR densities as high as $10^4\,\c-mmm$ are not very common. \OI\, and \CII\, \textit{ISO} observations of normal star-forming galaxies and some IR-bright sources confine the physical parameters of their PDRs to a range of $G_0\,\leq\,10^{4.5}$ and $10^2\,\lesssim\,$\nh$\,\lesssim\,10^4\,\c-mmm$ (\citealt{Malhotra2001}).
On the other hand, the \CIIno\, emission can be also saturated when $G_0\,>\,10^{1.5}$ provided that \nh$\,\lesssim\,10^3\,\c-mmm$. In this regime, the line is not sensitive to an increase of $G_0$ because the temperature of the gas is well above the excitation potential of the \CIIno\, transition.

It has also been suggested that in sources where \Go/\nh\, is high ($\gtrsim\,10^2\,\cmmm$)
the \CIIno\, line is a less efficient coolant of the ISM because of the following reason. As physical conditions become more extreme (higher \Go/\nh), dust particles progressively increase their positive charge (\citealt{Tielens1985}; \citealt{Malhotra1997}; \citealt{Negishi2001}). This reduces both the amount of photo-electrons released from dust grains that indirectly collisionally excite the gas, as well as the energy that they carry along after they are freed, since they are more strongly bounded. The net effect is the decreasing of the efficiency in the transformation of incident UV radiation into gas heating without an accompanied reduction of the dust emission (\citealt{Wolfire1990}; \citealt{Kaufman1999}; \citealt{Stacey2010}).

In a recent work, \cite{Gracia-Carpio2011} have shown that the deficits observed in several far-IR emission lines (\CII, \OI, \OIb, and \NIIa) could be explained by an increase of the average ionization parameter of the ISM, $<$\textit{U}$>$\symbolfootnote[1]{The ionization parameter is defined as \textit{U}$\,\equiv\,Q({\rm H})/4 \pi R^2 n_{\rm H} c$, where $Q({\rm H})$ is the number of hydrogen ionizing photons, \textit{R} is the distance of the ionizing source to the PDR, $n_{\rm H}$ is the atomic hydrogen density, and \textit{c} is the speed of light. If an average stellar population and size for the star-forming region is assumed, then \textit{U}$\,\propto\,$\Go/\nh.}.
In "dust bounded" star-forming regions the gas opacity is reduced within the \HII\, region due to the higher $<$\textit{U}$>$. As a consequence, a significant fraction of the UV radiation is eventually absorbed by large dust grains before being able to reach the neutral gas in the PDRs and ionize the PAH molecules (\citealt{Voit1992}; \citealt{GA2004}; \citealt{Abel2009}), causing a deficit of photo-electrons and hence the subsequent suppression of the \CIIno\, line with respect to the total far-IR dust emission.

Local luminous infrared galaxies (LIRGs: \LIR\,$=\,10^{11-12}$\,\Lsun) are a mixture of single galaxies, disk galaxy pairs, interacting systems and advanced mergers, exhibiting enhanced star formation rates, and a lower fraction of AGN compared to higher luminous galaxies.
A detailed study of the physical properties of low-redshift LIRGs is critical for our understanding of the cosmic evolution of galaxies and black holes since (1) IR-luminous galaxies comprise the bulk of the cosmic infrared background and dominate star-formation activity between $0.5\,<\,z\,<\,2$ (\citealt{Caputi2007}; \citealt{Magnelli2011}; \citealt{Murphy2011}; \citealt{Berta2011}) and (2) AGN activity may preferentially occur during episodes of enhanced nuclear star formation. Moreover, LIRGs are now assumed to be the local analogs of the IR-bright galaxy population at $z\,>\,1$. However, a comprehensive analysis of the most important far-IR cooling lines of the ISM in a complete sample of nearby LIRGs has not been possible until the advent of the \textit{Herschel Space Observatory} (\textit{Herschel} hereafter; \citealt{Pilbratt2010}) and, in particular, its Photodetector Array Camera and Spectrometer (PACS; \citealt{Poglitsch2010}).

In this work we present the first results obtained from \textit{Herschel}/PACS spectroscopic observations of a complete sample of far-IR selected local LIRGs that comprise the Great Observatories All-sky LIRG Survey (GOALS; \citealt{Armus2009}). Using this complete, flux-limited sample of local LIRGs, we are able for the first time to perform a systematic, statistically significant study of the far-IR cooling lines of star-forming galaxies covering a wide range of physical conditions: from isolated disks where star formation is spread across kpc scales to the most extreme environments present in late stage major mergers where most of the energy output of the system comes from its central kpc region. In particular, in this paper we focus on the \CII\, line and its relation with the dust emission in LIRGs. We make use of a broad set of mid-IR diagnostics based on \textit{Spitzer}/IRS spectroscopy, such as high ionization emission lines, silicate dust opacities, PAH equivalent widths (EW), dust luminosity concentrations, and mid-IR colors, to provide the context in which the observed \CIIno\, emission and \CIIno/FIR\, ratios are best explained. The paper is organized as follows: In \S\ref{s:sample} we present the LIRG sample and the observations. In \S\ref{s:datared} we describe the processing and analysis of the data. The results are presented in \S\ref{s:results}. In \S\ref{s:highz} we put in context our findings with recent results from intermediate and high redshift surveys started to be carried out by ALMA and in the future by CCAT. The summary of the results is given in \S\ref{s:summary}.

\section{Sample and Observations}\label{s:sample}

\subsection{The GOALS Sample}\label{ss:sample}

The Great Observatories All-sky LIRG Survey (GOALS; \citealt{Armus2009}) encompasses the complete sample of 202 LIRGs and ULIRGs contained in the \textit{IRAS} Revised Bright Galaxy Sample (RBGS; Sanders et al. 2003) which, in turn, is also a complete sample of 629 galaxies with \textit{IRAS} $S_{60\,\mu m}\,>\,5.24$\,Jy and Galactic latitudes $|b|\,>\,5\,^\circ$. There are 180 LIRGs and 22 ULIRGs in GOALS and their median redshift is z = 0.0215 (or $\sim\,95.2$\,Mpc), with the closest galaxy being at z = 0.0030 (15.9\,Mpc; NGC~2146) and the farthest at z = 0.0918 (400\,Mpc; IRAS~07251-0248). To date, there are many published and on-going works that have already exploited the potential of all the multi-wavelength data available for this sample including, among others, Galex UV \citep{Howell2010}, \textit{HST} optical and near-IR (\citealt{Haan2011}; Kim et al. 2013), and \textit{Chandra} X-ray \citep{Iwasawa2011} imaging, as well as \textit{Spitzer}/IRS mid-IR spectroscopy (\citealt{DS2010b, DS2011}; \citealt{Petric2011}; Stierwalt et al. 2013a,b; Inami et al. 2013), as well as a number of ground-based observatories (VLA, CARMA, etc.) and soon ALMA.

The RBGS, and therefore the GOALS sample, were defined based on \textit{IRAS} observations. However, the higher angular resolution achieved by \textit{Spitzer} allowed us to spatially disentangle galaxies that belong to the same LIRG system into separate components. From the 291 \textit{individual} galaxies in GOALS, not all have \textit{Herschel} observations. In systems with two or more galactic nuclei, minor companions with MIPS\,24$\,\mu$m flux density ratios smaller than 1:5 with respect to the brightest galaxy were not requested since their contribution to the total IR luminosity of the system is small.
Because the angular resolution of \textit{Spitzer} decreases with wavelength, it was not possible to obtain individual MIPS 24, 70 and 160$\,\mu$m measurements for all GOALS galaxies, and therefore to derive uniform IR luminosities for them using \textit{Spitzer} data only. Instead, to calculate the individual, spatially-integrated \LIR\, of LIRGs belonging to a system of two or more galaxies, we distributed the $L_{\rm IR}^{\rm 8-1000\,\mu m}$ of the system as measured by \textit{IRAS} (using the prescription given in \citealt{Sanders1996}) proportionally to the individual MIPS\,70$\,\mu$m flux density of each component when available, or to their MIPS\,24$\,\mu$m otherwise\footnote{There are two systems for which no individual MIPS\,24$\,\mu$m fluxes could be obtained. In these cases their IRAC\,8$\,\mu$m emission was used for scaling the \LIR. These LIRGs are: MCG+02-20-003 and VV250a.}. We will use this measurements of \LIR\, in \S\ref{s:highz}

\subsection{Herschel/PACS Observations}\label{ss:pacsobs}

We have obtained far-IR spectroscopic observations for 153 LIRG systems of the GOALS sample using the Integral Field Spectrometer (IFS) of the PACS instrument on board \textit{Herschel}. The data were collected as part of an OT1 program (OT1\_larmus\_1; P.I.: L. Armus) awarded with more than 165 hours of observing time. In this work will focus mainly on the analysis and interpretation of the \CIIno\, observations of our galaxy sample.
PACS range spectroscopy of the \CII\, fine-structure emission line was obtained for 163 individual sources. Our observations were complemented with the inclusion of the remaining LIRGs in the GOALS sample for which \CIIno\, observations are publicly available in the archive (as of October 2012) from various \textit{Herschel} projects. The main programs from which these data were gathered are: KPGT\_esturm\_1 (P.I.: E. Sturm), KPOT\_pvanderw\_1 (P.I.: P. van der Werf), and OT1\_dweedman\_1 (P.I.: D. Weedman). The total number of LIRG systems for which there are \CIIno\, data is 200 (IRASF08339+6517 and IRASF09111-1007 were not observed). However, because some LIRGs are actually systems of galaxies (see above), the number of observed galaxies was 241.

The IFS on PACS is able to perform simultaneous spectroscopy in the $51-73$ or $70-105\,\mu$m (3rd and 2nd orders, respectively; "blue" camera) and the $102-210\,\mu$m (1st order; "red" camera) ranges.
The IFU is composed by a 5\,$\times$\,5 array of individual detectors (spaxels) each of one with a field of view (FoV) of $\sim\,$9.4\arcsec, for a total of 47\arcsec\,$\times$\,47\arcsec. The physical size of the PACS FoV at the median distance of our LIRG sample is $\sim$\,20\,kpc on a side. The number of spectral elements in each pixel is 16, which are rearranged together via an image slicer over two 16\,$\times$\,25 Ge:Ga detector arrays (blue and red cameras).

Our Astronomical Observation Requests (AORs) were consistently constructed using the "Range" spectroscopy template, which allows the user to define a specific wavelength range for the desired observations. Our selected range was slightly larger than that provided by default for the "Line" mode. This was necessary (1) to obtain parallel observations of the wide \OH\, absorption feature using the blue camera when observing the \CII\, line, and (2) to ensure that the targeted emission lines have a uniform signal-to-noise ratio across their spectral profiles even if they are to be broader than a few hundred \kms. The high sampling density mode scan, useful to have sub-spectral resolution information of the lines (see below), was employed. While we requested line maps for some LIRGs of the sample (from two to a few raster positions depending on the target), pointed (one single raster) chop-nod observations were taken for the majority of galaxies. For those galaxies with maps, only one raster position was used to obtain the line fluxes used in this work. The chopper throw varied from small to large depending on the source. Spectroscopy of the LIRGs included in GOALS but observed by other programs in \CIIno\, was not always obtained using the "Range" mode but some of them were observed using "LineScan" spectroscopy. The S/N of the data varies not only from galaxy to galaxy but also depending on the emission line considered. We provide uncertainties for all quantities used across the analysis presented here that are based on the individual spectrum of each line, therefore reflecting the errors associated with $-$and measured directly on$-$ the data.

\subsection{Spitzer/IRS Spectroscopy}\label{ss:irsobs}

As part of the \textit{Spitzer} GOALS legacy, all galaxies observed with \textit{Herschel}/PACS have available \textit{Spitzer}/IRS low resolution (R\,$\sim\,60-120$) slit spectroscopy (SL module: $5.5-14.5\,\mu$m, and LL module: $14-38\,\mu$m). The 244 IRS spectra were extracted using the standard extraction aperture and point source calibration mode in SPICE. The projected angular sizes of the apertures on the sky are 3.7\arcsec\,$\times$\,12\arcsec\, at the average wavelength of 10$\,\mu$m in SL and 10.6\arcsec\,$\times$\,35\arcsec\, at the average wavelength of 26$\,\mu$m in LL. Thus, the area covered by the SL aperture is approximately equivalent (within a factor of $\sim\,2$) to that of an individual spaxel of the IFS in PACS, and so is that of the LL aperture to a 3\,$\times$\,3 spaxel box. The observables derived from the IRS data that we use in this work are the strength of the 9.7$\,\mu$m silicate feature, \SSi, and the EW of the \PAHc, which were presented in Stierwalt et al. (2013a). We refer the reader to this work for further details about the reduction, extraction, calibration, and analysis of the spectra.

\section{IFS/PACS Data Reduction and Analysis}\label{s:datared}

\subsection{Data Processing}\label{ss:dataproc}

The \textit{Herschel} Interactive Processing Environment (HIPE; v8.0) application was used to retrieve the raw data from the \textit{Herschel} Science Archive (HSA\footnote{http://herschel.esac.esa.int/Science\_Archive.shtml}) as well as to process them. We used the script for "LineScan" observations (also valid for "Range" mode) included within HIPE to reduce our spectra. We processed the data from level 0 up to level 2 using the following steps:  Flag and reject saturated data, perform initial calibrations, flag and reject "glitches", compute the differential signal of each on-off pair of data-points for each chopper cycle, calculate the relative spectral response function, divide by the response, convert frames to PACS cubes, and correct for flat-fielding (this extra step is included in v8.0 of HIPE and later versions, and helps to improve the accuracy of the continuum level). Next, for each camera (red or blue), HIPE builds the wavelength grid, for which we chose a final rebinning with an \textit{oversample}\,=\,2, and an \textit{upsample}\,=\,3 that corresponds to a Nyquist sampling.
The spectral resolution achieved at the position of the \CII\, line was derived directly from the data and is $\sim\,$235\,\kms. The final steps are: flag and reject remaining outliers, rebin all selected cubes on consistent wavelength grids and, finally, average the nod-A and nod-B rebinned cubes (all cubes at the same raster position are averaged). This is the final science-grade product currently possible for single raster observations. From this point on, the analysis of the spectra was performed using in-house developed IDL routines.


\subsection{Data Analysis}\label{ss:dataanalysis}

To obtain the \CIIno\, flux of a particular source we use an iterative procedure to find the line and measure its basic parameters. First, we fit a linear function to the continuum emission, which is evaluated at the edges of the spectrum, masking the central 60\,\% of spectral elements (where the line is expected to be detected) and without using the first and final 10\,\%, where the noise is large due to the poor sampling of the scanning.
Then, we fit a Gaussian function to the continuum-subtracted spectrum and calculate its parameters. We define a line as not detected when the peak of the Gaussian is below 2.5$\,\times$ the standard deviation of the continuum, as measured in the previous step. On the other hand, if the line is found, we return to the original, total spectrum and fit again the continuum using this time a wavelength range determined by the two portions of the spectrum adjacent to the line located beyond $\pm\,3\,\sigma$ from its center (where $\sigma$ is the width of the fitted Gaussian) and the following $\pm\,$15\,\% of spectral elements. We then subtract this continuum from the total spectrum and fit the line again. The new parameters of the Gaussian are compared with the previous ones. This process is repeated until the location, sigma and intensity of the line converge with an accuracy of 1\,\%, or when reaching 10 iterations.
Due to the merger-driven nature of many LIRGs, their gas kinematics are extremely complicated and, as a consequence, the emission lines of several sources present asymmetries and double peaks in their profiles. However, despite the fact that the width determined by the fit is not an accurate representation of the real shape of the line, it can be used as a first order approximation for its broadness. Therefore, instead of using the parameters of the Gaussian to derive the flux of the line, we decided to integrate directly over the final continuum-subtracted spectrum within the $\pm\,3\,\sigma$ region around the central position of the line. The associated uncertainty is calculated as the standard deviation of the latest fitted continuum, integrated over the same wavelength range as the line. Absolute photometric uncertainties due to changes in the PACS calibration products are not taken into account (the version used in this work was PACS\_CAL\_32\_0)\footnote{http://herschel.esac.esa.int/twiki/bin/view/Public/PacsCalTreeHistory}.

We obtained the line fluxes for our LIRGs from the spectra extracted from the spaxel at which the \CIIno\, line + continuum emission of each galaxy peaks within the PACS FoV.
The \textit{Spitzer}/IRS and \textit{Herschel} pointings usually coincide within $\lesssim\,2\arcsec$.
There are a few targets for which the IRS pointing is located more than half a spaxel away from that of PACS. In these cases, we decided to obtain the nuclear line flux of the galaxy by averaging the spaxels closest to the coordinates of the IRS pointing. These values are used only when PACS and IRS measurements are compared directly in the same plot. There is one additional LIRG system, IRAS03582+6012, for which the PACS pointing exactly felt in the middle of two galaxies separated by only 5\arcsec. This LIRG is not used in the comparisons of the \CIIno\, emission to the IRS data since the two individual sources cannot be disentangled.

As mentioned in \S\ref{ss:irsobs}, the angular size of a PACS spaxel is roughly similar (within a factor of 2) to that of the aperture used to extract the \textit{Spizer}/IRS spectra of our galaxies. Because the PACS beam is under-sampled at 160$\,\mu$m (FWHM$\,\sim\,12\arcsec$ compared with the $9.6\arcsec$ size of the PACS spaxels), and most of the sources in the sample are unresolved at 24$\,\mu$m in our MIPS images (which have a similar angular resolution as PACS at $\sim\,80\,\mu$m), an aperture correction has to be performed to the spectra extracted from the emission-peak spaxel of each galaxy to obtain their total nuclear fluxes. This was the same procedure employed to obtain the mid-IR IRS spectra of our LIRGs.
The nominal, wavelength-dependent aperture correction function provided by HIPE v8.0 works optimally when the source is exactly positioned at the center of a given spaxel. However, in some occasions the pointing of \textit{Herschel} is not accurate enough to achieve this and the target can be slightly misplaced $\lesssim\,3\,\arcsec$ (up to 1/3 of a spaxel) from the center. In these cases, the flux of the line might be underestimated.
We explored whether this effect could be corrected by measuring the position of the source within the spaxel. However, some LIRGs in our sample show low surface-brightness extended emission, either because of their proximity and/or merger nature, or simply because the gas and dust emission are spatially decoupled. This, combined with the spatial sub-sampling of the PACS/IFS detector and the poor S/N of some sources prevented us from obtaining an accurate measurement of the spatial position and angular width of the \CIIno\, emission and therefore from obtaining a more refined aperture correction. Thus, we performed only the nominal aperture correction provided by HIPE.

The \textit{IRAS} far-IR fluxes used throughout this paper were calculated as ${\rm FIR}\,=\,1.26\,\times\,10^{-14}\,(2.58\,S_{\rm 60\mu m}\,+\,S_{\rm 100\mu m})$ [\wm], with $S_\nu$ in [Jy]. The far-IR luminosities, \LFIR, were defined as $4\pi\,D_{\rm L}^2\,{\rm FIR}\,$ [\Lsun]. The luminosity distances, $D_{\rm L}$, were taken from \cite{Armus2009}. This definition of the FIR accounts for the flux emitted within the $42.5-122.5\,\mu$m wavelength range as originally defined in \cite{Helou1988}.
The far-IR fluxes and luminosities of galaxies were then matched to the aperture with which the nuclear \CIIno\, flux was extracted (see above) by
scaling the integrated \textit{IRAS} far-IR flux of the LIRG system with the ratio of the continuum flux density of each individual galaxy evaluated at 63$\,\mu$m in the PACS spectrum (extracted at the same position and with the same aperture as the \CIIno\, line) to the total \textit{IRAS} 60$\,\mu$m flux density of the system.

In Table~\ref{t:sample} we present the \CII\, flux, the \CIIno/FIR ratio, and the continuum flux densities at 63 and 158$\,\mu$m for all the galaxies in our sample. Future updates of the data in this table processed with newer versions of HIPE and PACS calibration files will be available at the GOALS webpage: http://goals.ipac.caltech.edu.

\begin{deluxetable*}{lccccccc}
\tabletypesize{\scriptsize}
\tablewidth{0pc}
\tablecaption{\scriptsize \textit{Herschel}/PACS measurements for the GOALS sample}
\tablehead{\colhead{Galaxy} & \colhead{R.A.} & \colhead{Dec.} & \colhead{Dist.}& \colhead{\CII} & \colhead{\CIIno/FIR} & \colhead{$S_\nu$63$\mu$m cont.} & \colhead{$S_\nu$158$\mu$m cont.} \\
\colhead{name} & \colhead{[hh:mm:ss]} & \colhead{[dd:mm:ss]} & \colhead{[Mpc]} & \colhead{[$\times\,10^{-15}\,$\wm]} & \colhead{[$\times\,10^{-3}$]} & \colhead{[Jy]} & \colhead{[Jy]} \\
\colhead{(1)} & \colhead{(2)} & \colhead{(3)} & \colhead{(4)} & \colhead{(5)} & \colhead{(6)} & \colhead{(7)} & \colhead{(8)}}
\startdata 
NGC0023              &     00h09m53.4s &  +25\deg 55m26s &    65.2 &   1.385 $\pm$  0.022 &    4.13 $\pm$   0.09 &    6.17 $\pm$   0.08 &    6.85 $\pm$   0.08 \\
NGC0034              &     00h11m06.5s &  $-$12\deg 06m26s &    84.1 &   0.624 $\pm$  0.018 &    0.85 $\pm$   0.03 &   16.39 $\pm$   0.13 &    9.02 $\pm$   0.06 \\
Arp256               &     00h18m50.9s &  $-$10\deg 22m36s &   117.5 &   0.967 $\pm$  0.020 &    2.96 $\pm$   0.07 &    6.70 $\pm$   0.08 &    4.42 $\pm$   0.09 \\
\dots & \dots & \dots & \dots & \dots & \dots & \dots & \dots
\enddata
\tablecomments{\scriptsize (1) Galaxy name; (2)$-$(3) Right Ascension and Declination (J2000) of the position from which the \textit{Herschel}/PACS spectrum was extracted (\S\ref{ss:dataanalysis}); (4) Distance to the galaxies taken from \cite{Armus2009}; (5) \CII\, flux as measured from the spaxel at which the \CIIno\, line + continuum emission of the galaxy peaks within the PACS FoV; that is, within an effective aperture of $\sim\,9.4\arcsec\,\times\,9.4\,\arcsec$ (\S\ref{ss:dataanalysis}); (6) \CIIno\, to far-IR flux ratio, where the far-IR fluxes have been scaled to match the aperture of the \CIIno\, measurements; (7) Continuum flux density at 63$\mu$m under the \OIno\, line extracted at the same position and with the same aperture size as (5); (8) Same as (7) but for the continuum at 158$\mu$m under the \CIIno\, line.\\
There are 11 galaxies for which the \textit{Spitzer}/IRS pointing is located $>\,4.7\arcsec\,$ from the position of the \CIIno +continuum peak. For these, we include an extra entry in the table with the \CIIno\, flux and continuum measurements obtained at the position of the IRS slit. They are marked with asterisks next to the names.\\
The complete table is available in the electronic edition of the paper. Future updates of the data in this table processed with newer versions of HIPE and PACS calibration ﬁles will be available at the GOALS webpage: http://goals.ipac.caltech.edu}\label{t:sample}
\end{deluxetable*}

\section{Results and Discusion}\label{s:results}

\subsection{The \CIIno/FIR\, Ratio: Dust Heating and Cooling}\label{s:dusttemp}

\begin{figure}[t!]
\vspace{.25cm}
\epsscale{1.15}
\plotone{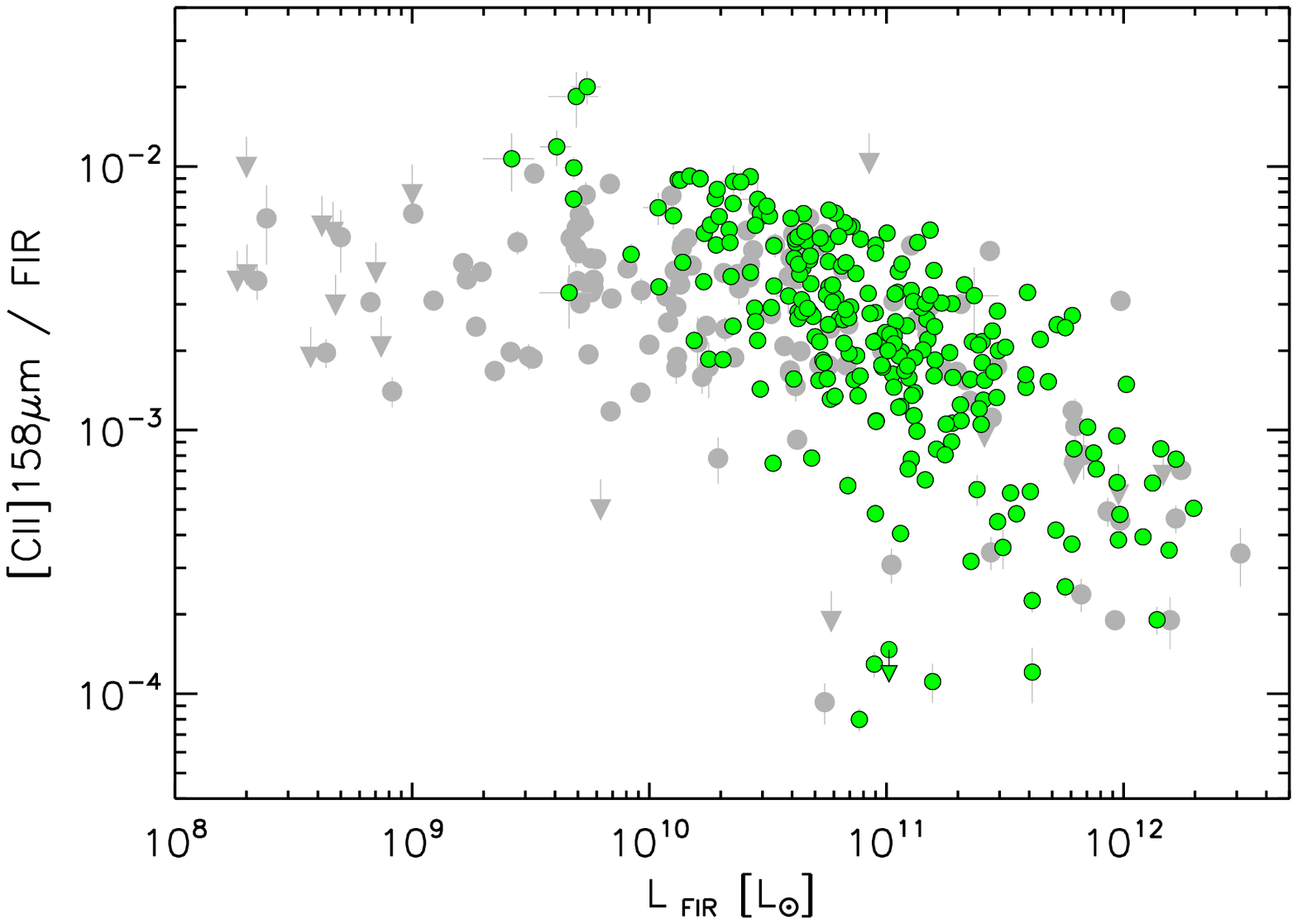}
\vspace{.25cm}
\caption{\footnotesize The ratio of \CII\, to far-IR flux as a function of the far-IR luminosity for individual galaxies in the GOALS sample (green circles) and for unresolved galaxies observed with \textit{ISO} (gray circles and limits) obtained from the compilation of \cite{Brauher2008} located at $z\,>\,0.003$, similar to the distance range covered by our LIRGs. The \LFIR\, of galaxies was calculated as explained at the end of \S\ref{ss:dataanalysis} and covers the $42.5-122.5\,\mu$m wavelength range as defined in \cite{Helou1988}.}\label{f:ciifirvsfir}
\vspace{.5cm}
\end{figure}

The far-IR fine structure line emission in normal star-forming galaxies as well as in the extreme environments hosted by ULIRGs has been extensively studied for the past two decades. A number of works based on \textit{ISO} data already suggested that the relative contribution of the \CII\, line to the cooling of the ISM in PDRs compared to that of large dust grains, as gauged by the far-IR emission, diminishes as galaxies are more IR luminous (\citealt{Malhotra1997}; \citealt{Luhman1998}; \citealt{Brauher2008}). Figure~\ref{f:ciifirvsfir} display the classical plot of the \CIIno/FIR\, ratio as a function of the far-IR luminosity for our LIRG sample. In addition, we also show for reference those galaxies observed with \textit{ISO} compiled by \cite{Brauher2008} that are classified as unresolved and located at redshifts $z\,>\,0.003$, similar to the distance range covered by GOALS. As we can see, our \textit{Herschel} data confirm the trend seen with \textit{ISO} by which galaxies with \LFIR\,$\gtrsim\,10^{11}\,\Lsun$ show a significant decrease of the \CIIno/FIR\, ratio. GOALS densely populates this critical part of phase-space providing a large sample of galaxies with which to explore the physical conditions behind the drop in \CIIno\, emission among LIRGs. For the 32 galaxies with measurements obtained with both telescopes, the higher angular resolution \textit{Herschel} observations of the nuclei of LIRGs are able to recover an average of $\sim\,$87\% of the total \CIIno\, flux measured by \textit{ISO}.


\subsubsection{The Average Dust Temperature of LIRGs}\label{ss:warm}


\begin{figure}[t!]
\vspace{.25cm}
\epsscale{1.15}
\plotone{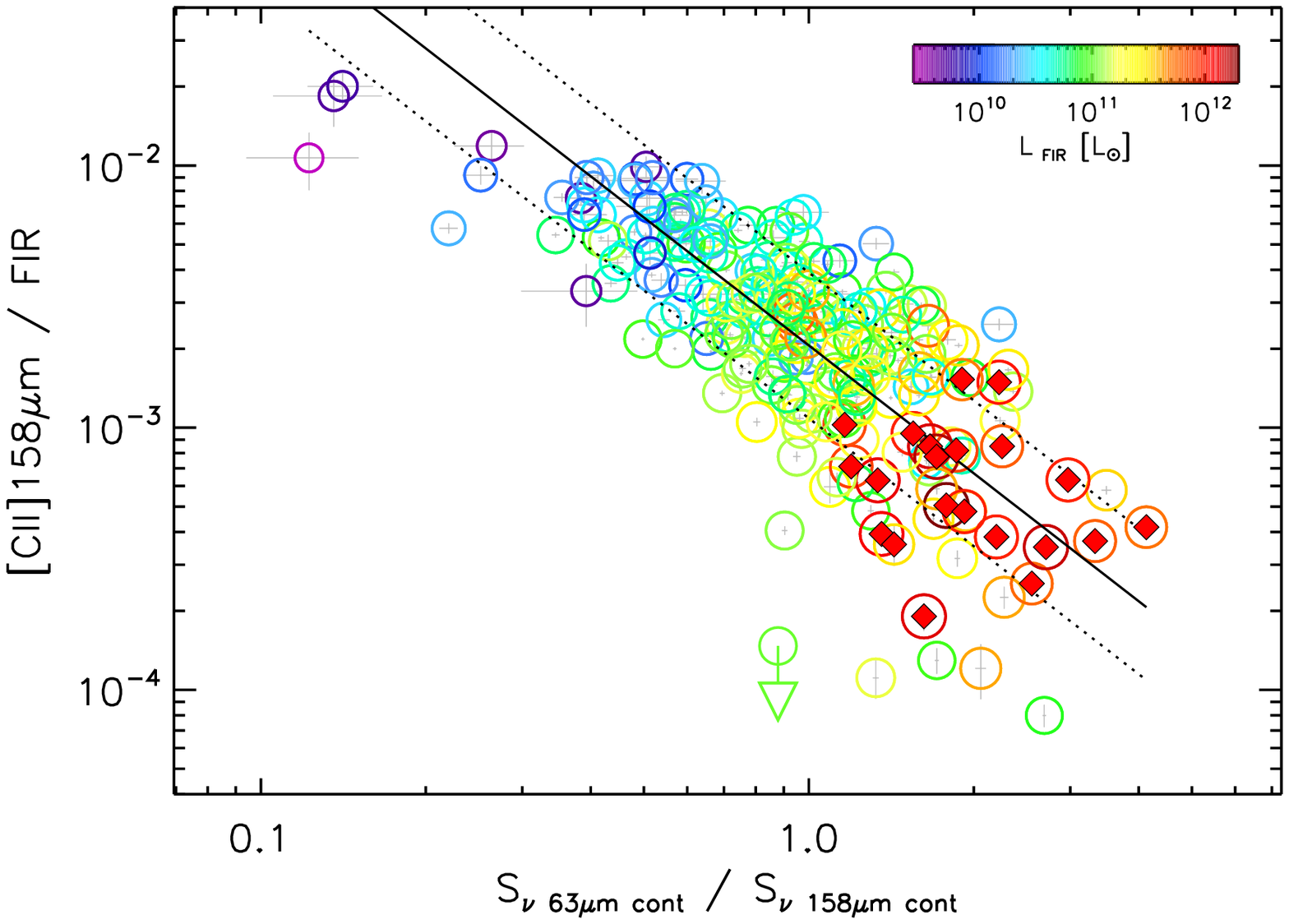}
\vspace{.5cm}
\plotone{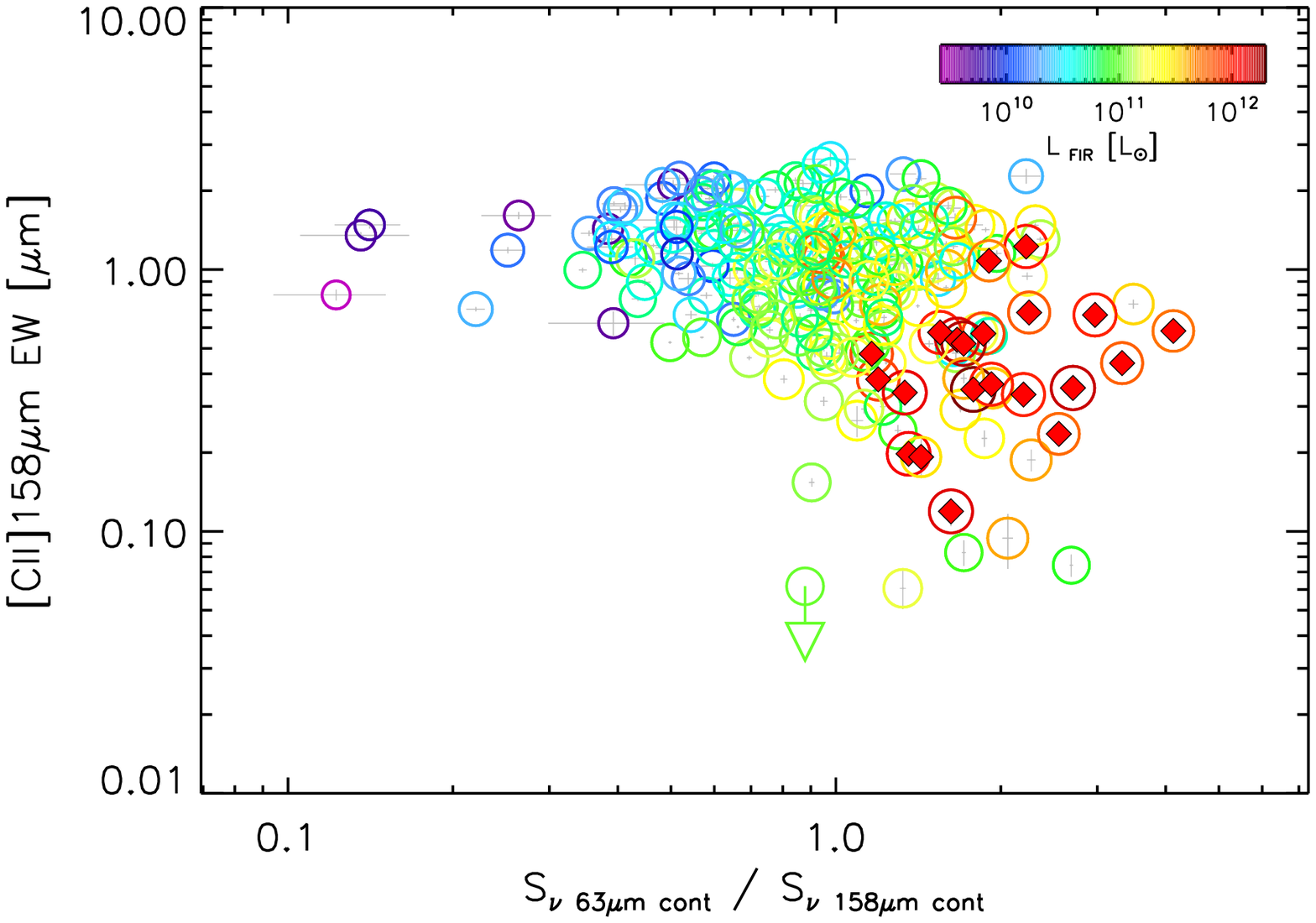}
\vspace{.25cm}
\caption{\footnotesize The ratio of \CII\, to far-IR flux (upper panel) and \CII\, EW (bottom panel) as a function of the $S_\nu\,$63\,$\mu$m/$S_\nu\,$158\,$\mu$m continuum flux density ratio for individual galaxies in the GOALS sample. Circles of different colors indicate the \LFIR\, of galaxies (see color bar), which is defined as explained in \S\ref{ss:dataanalysis}. Red diamonds mark galaxies with $\LIR\,\geq\,10^{12}\,\Lsun$, ULIRGs. These two plots show that the decrease of the \CIIno/FIR\, ratio with warmer far-IR colors seen in our LIRGs is primarily caused by a significant decrease of gas heating efficiency and an increase of warm dust emission. The solid line in the upper panel corresponds to a linear fit of the data in log-log space. The parameters of the fit are given in Eq.~(1). The dotted lines are the $\pm\,1\,\sigma$ uncertainty. 
}\label{f:ciifirvsf63f158}
\vspace{.5cm}
\end{figure}

Figure~\ref{f:ciifirvsf63f158} (upper panel) shows the \CII/FIR\, ratio for the GOALS sample as a function of the far-IR PACS $S_\nu\,$63\,$\mu$m/$S_\nu\,$158\,$\mu$m continuum flux density ratio. We chose to use this PACS-based far-IR color in the x-axis instead of the more common \textit{IRAS} 60/100\,$\mu$m color mainly because of two main reasons: (1) this way we are able plot data from individual galaxies instead of being constrained by the spatial resolution of \textit{IRAS}, which would force us to show only blended sources; (2) by using the 63/158\,$\mu$m ratio we are probing a larger range of dust temperatures within the starburst ($T\,\sim\,50\,\rightarrow\,20\,$K); with the colder component probably arising from regions located far from the ionized gas-phase, and closer to the PDRs where the \CIIno\, emission originates. For reference, we show the relation between the PACS 63/158\,$\mu$m and \textit{IRAS} 60/100\,$\mu$m colors in the Appendix. The ULIRGs in the GOALS sample (red diamonds) have a median \CIIno/FIR\,=\,$6.3\,\times\,10^{-4}$, a mean of $6.6\,$($\pm\,0.8$)$\,\times\,10^{-4}$, and a standard deviation of the distribution of $3.7\,\times\,10^{-4}$. LIRGs span two orders of magnitude in \CIIno/FIR, from $\sim\,10^{-2}$ to $\sim\,10^{-4}$, with a mean of $3.4\,\times\,10^{-3}$ and a median of $2.6\,\times\,10^{-3}$. The \LCIIno\, ranges from $\sim\,10^7$ to 2$\,\times\,10^9\,$\Lsun.

Our results are consistent with \textit{ISO} observations of a sample of normal and moderate IR-luminous galaxies presented in \cite{Malhotra1997} and further analyzed in \cite{Helou2001}. The GOALS sample, though, populates a warmer far-IR color regime. Despite the increase in dispersion at 63/158\,$\mu$m$\,\gtrsim\,1.25$ or \CII/FIR\,$\lesssim\,10^{-3}$ (basically in the ULIRG domain), the fact that we find the same tight trend independently of the range of IR luminosities covered by the two samples suggests that the main observable linked to the variation of the \CIIno/FIR\, ratio is
the average temperature of the dust (\Tdust) in galaxies.

This interpretation agrees with the last physical scenario described in the Introduction, in which an increase of the ionization parameter, $<$\textit{U}$>$, would cause the far-UV radiation from the youngest stars to be less efficient in heating the gas in those galaxies. At the same time, dust grains would be on average at higher temperatures due to the larger number of ionizing photons per dust particle available in the outer layers of the \HII\, regions, close to the PDRs. Indeed, the presence of dust within \HII\, regions has been recently observed in several star-forming regions in our Galaxy (\citealt{Paladini2012}). Both effects combined can explain the wide range of \CII/FIR\, ratios and far-IR colors we observe in the most warm LIRGs. Variations in \nh, though, could be responsible for the dispersion in \CIIno/FIR\, seen at a given 63/158\,$\mu$m ratio.

To further support these findings, the bottom panel of Figure~\ref{f:ciifirvsf63f158} shows that the ratio of \CII\, flux to the monochromatic continuum at $\sim\,$158$\,\mu$m under the line (the \CIIno\, EW) of the warmest galaxies is only a factor of $\sim\,4$ lower than the average \CIIno\, EW displayed by colder sources at 63/158\,$\mu$m\,$\lesssim\,1$.
This implies that the decrease of the \CIIno/FIR\, ratio seen in our LIRGs is primarily caused by a significant increase in warm dust emission (peaking at $\lambda\,\lesssim\,60-100\,\mu$m), most likely associated with the youngest stars, that is not followed by a proportional enhancement of the \CIIno\, emission line.

The best fit to the data in Figure~\ref{f:ciifirvsf63f158} (upper panel) yields the following parameters:

\begin{equation}
log(\frac{\CIIno}{\rm FIR})\,=\,-2.68(\pm 0.02) - 1.61(\pm 0.09)\,log(\frac{S_{63\mu m}}{S_{158\mu m}})
\end{equation}\label{e:defvsfircolorlin}

\noindent
with a dispersion of 0.28\,dex. We note that the \CIIno/FIR\, ratios predicted by the fitted relation for sources with far-IR colors 63/158\,$\mu$m$\,\lesssim\,0.4$ are probably overestimated, as it is already know that galaxies showing such cool \Tdust\, have typical \CIIno/FIR\,$\,\sim\,10^{-2}$ (e.g., \citealt{Malhotra2001}).

\subsubsection{The Link Between Mid-IR Dust Obscuration and Far-IR Re-emission}\label{ss:siabs}

\begin{figure}[t!]
\vspace{.25cm}
\epsscale{1.15}
\plotone{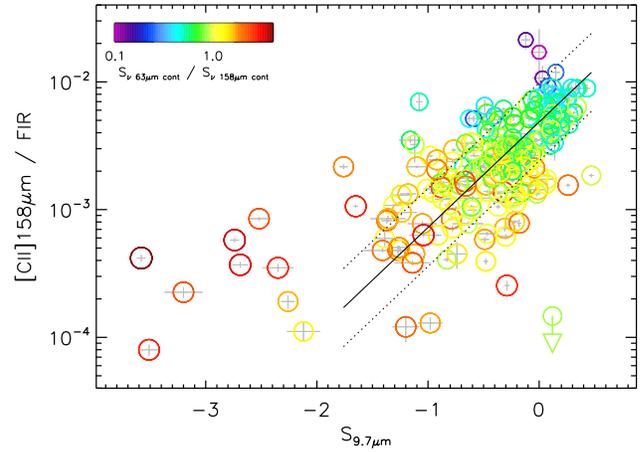}
\vspace{.25cm}
\caption{\footnotesize \CII/FIR\, ratio for individual galaxies in the GOALS sample as a function of the strength of the 9.7\,$\mu$m silicate absorption feature, \SSi, measured with \textit{Spitzer}/IRS. Galaxies are color-coded as a function of the $S_\nu\,$63\,$\mu$m/$S_\nu\,$158\,$\mu$m ratio, with values ranging from 0.10 to 4.15 (see Figure~\ref{f:ciifirvsf63f158}). The solid line represents an outlier-resistant fit to the bulk of the star-forming LIRG population with \SSi\,$\geq\,-2$. The dotted lines are the $\pm\,1\,\sigma$ uncertainty.}\label{f:ciifirvssiabs}
\vspace{.5cm}
\end{figure}

The strength of the 9.7\,$\mu$m silicate feature is defined as $\SSi\,\equiv\,ln(f_{\lambda_P}^{obs}/f_{\lambda_P}^{cont})$; with $f_{\lambda_P}^{cont}$ and $f_{\lambda_P}^{obs}$ being the un-obscured and observed continuum flux density measured in the mid-IR IRS spectra of our LIRGs and evaluated at the peak of the feature, $\lambda_P$, normally at 9.7$\,\mu$m (see Stierwalt et al. 2013a for details on how it was calculated in our sample). Negative values indicate absorption, while positive ones indicate emission. By definition, \SSi\, measures the apparent optical depth towards the warm, mid-IR emitting dust. Figure~\ref{f:ciifirvssiabs} shows that there is a clear trend ($r\,=\,0.65,\,p_r=0; \kappa\,=0.47$) for LIRGs with stronger (more negative) \SSi\, to display smaller \CIIno/FIR\, ratios, implying that the dust responsible for the mid-IR absorption is also accountable for the far-IR emission. The formal fit (solid line) can be expressed as:

\begin{equation}
log(\frac{\CIIno}{\rm FIR})\,=\,-2.32(\pm 0.02) + 0.83(\pm 0.05)\,\SSi
\end{equation}\label{e:defvslsd}

\noindent
with a dispersion in the y-axis of 0.30\,dex.


Within the context described in the previous section, the contrast between the inner layer of dust that is being heated by the ionizing radiation to $T\,\gtrsim\,50$\,K and that of the cold dust at $T\,\lesssim\,$20\,K emitting at $\lambda\,\gtrsim\,150\,\mu$m would create both: (1) the silicate absorption seen at 9.7\,$\mu$m due to the larger temperature gradient between the two dust components and (2) the increasingly higher 63/158\,$\mu$m ratios seen in Figure~\ref{f:ciifirvsf63f158} due to the progressively larger amount of dust mass that is being heated to higher temperatures.
This scenario is consistent with the physical properties of the ISM found in the extreme environments of ULIRGs, in which the fraction of total dust luminosity contributed by the diffuse ISM decreases significantly, and the emission from dust at $T\,\sim\,50-60$\,K arising from optically-thick "birth clouds" (with ages $\lesssim\,10^{7-8}\,$Myr) accounts for $\gtrsim\,$80\% of their IR energy output (\citealt{daCunha2010}). Furthermore, our findings are also in agreement with recent results showing that the increase of the silicate optical depth in LIRGs is related with the flattening of their radio spectral index (1.4 to 8.44\,GHz) due to an increase of free-free absorption, suggesting that the dust obscuration must largely be originated in the vicinity and/or within the starburst region (Murphy et al. 2013).

There are a few galaxies that do not follow the correlation fitted in Figure~\ref{f:ciifirvssiabs}, showing very large silicate strengths (\SSi\,$<\,-2$)
and small \CIIno/FIR\, ratios ($<\,10^{-3}$) typical of ULIRGs or, in general, warm galaxies (see color coding or Figure~\ref{f:ciifirvsf63f158}). We would like to note that the trend found for the majority of our sample only reaches \SSi\, values up to around $-1.5$ which, interestingly, is only slightly larger than the apparent optical depth limit that a obscuring clumpy medium can explain (\citealt{Nenkova2008b}). Larger (more negative) values of the silicate strength can only be achieved by a geometrically thick (smooth) distribution of cold dust, suggesting that the extra dust absorption seen in these few galaxies may not be related with the star-forming region from where the \CIIno\, and far-IR emissions arise.

But then, what is the origin of this excess of obscuration? One possibility is that it is caused by foreground cold dust not associated with the starburst. This has been seen in some heavily obscured Compton-thick AGNs, where most of the deep silicate absorption measured in these objects seems to originate from dust located in the host galaxy (\citealt{Goulding2012}; Gonzalez-Martin et al. 2012). Alternatively, the presence of an extremely warm source (different from the star-forming region(s) that are producing the far-IR and \CIIno\, emission) could contribute with additional emission of hot dust (T$\,\gtrsim\,150\,$K) to the mid-IR. If at the same time this source is deeply buried (optically thick) and embedded in layers of progressively colder dust (geometrically thick), it could produce a cumulative absorption that we would measure via the strength of the silicate feature while still contributing to the emission outside of it (see \citealt{Levenson2007}; \citealt{Sirocky2008}).

While both explanations are plausible, the second is favored by the fact that these extremely obscured galaxies show MIPS 24/70$\,\mu$m ratios very similar, or even slightly higher than those found for the rest of the LIRGs in the sample. If foreground cold dust was the responsible for the excess of obscuration, we would expect these galaxies to show abnormally low 24$\,\mu$m luminosities with respect to the far-IR. We find that this is not the case, in agreement with recent results based on radio observations of a sub-sample of LIRGs in GOALS (Murphy et al. 2013). Furthermore, the existence of an additional hot and obscured dust component in these LIRGs is also consistent with the results presented in Stierwalt et al. (2013a), where it is shown that there is a trend for LIRGs with moderate silicate strengths ($\SSi\,\gtrsim\,-1.5$) to show higher $S_\nu\,$30\,$\mu$m/$S_\nu\,$15\,$\mu$m ratios as the \SSi\, becomes stronger (more negative). That is, more obscured LIRGs have increasingly larger fluxes at 30\,$\mu$m, in agreement with our findings in the previous section. However, galaxies showing the most extreme silicate strengths ($\SSi\,<-1.5$) do not have proportionally higher 30/15\,$\mu$m ratios. On the contrary, they show ratios similar to those of warm LIRGs with mild silicate strengths (or even lower than expected given their extreme \SSi), supporting the idea that in these particular galaxies the dust producing this additional absorption and excess of mid-IR emission ($\lambda\,\lesssim\,20\,\mu$m) represents a component of the overall nuclear starburst activity different than the star-forming regions that drive the far-IR cooling.

\subsubsection{The Compactness of the Mid-IR Emitting Region}\label{ss:compact}

The compactness of the starburst region of a galaxy has been proven to be related to many of its other physical properties (\citealt{Wang1992}). For example, all ULIRGs in the GOALS sample have very small mid-IR emitting regions, with sizes (measured FWHMs) $<\,1.5\,$kpc (\citealt{DS2010b}). LIRGs, on the other hand, span a large range in sizes as well as in how much of their mid-IR emission is extended. The later property is parametrized in \cite{DS2010b} by the fraction of extended emission, FEE$_\lambda$, which measures the fraction of light emitted by a galaxy that is contained outside of its unresolved component at a given wavelength $\lambda$. The complementary quantity $1-$FEE$_\lambda$ measures how compact the source is, which in turn is proportional to its luminosity surface density, $\Sigma$. We note that in this paper we use the word compactness as an equivalent to light concentration, i.e., as a measurement of the amount of energy per unit area produced by a source, and not as an absolute measurement of its size.

\begin{figure}[t!]
\vspace{.25cm}
\epsscale{1.15}
\plotone{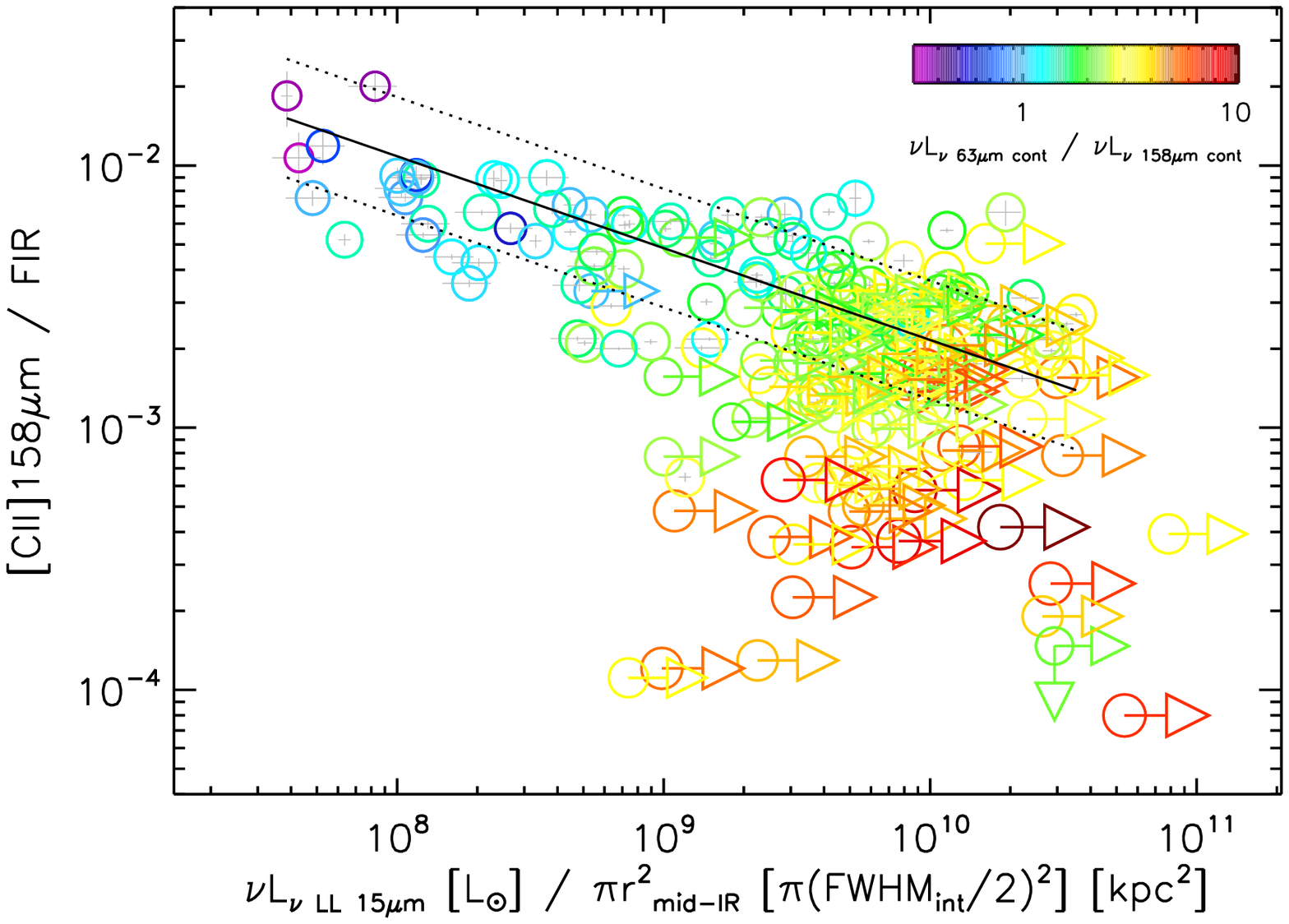}
\vspace{.5cm}
\plotone{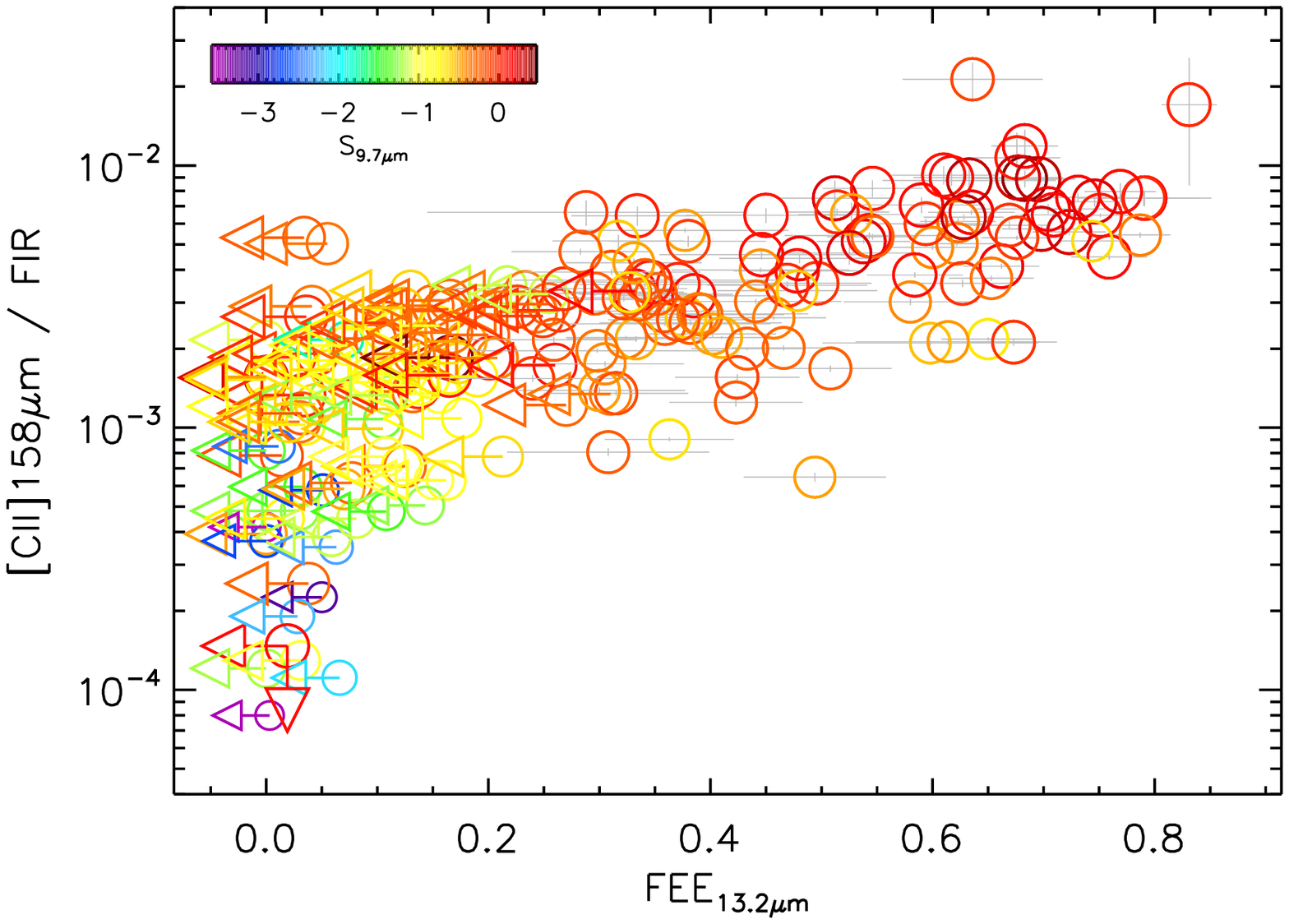}
\vspace{.25cm}
\caption{\footnotesize \CII/FIR\, ratio as a function of the luminosity surface density at 15$\,\mu$m, $\Sigma_{15\,\mu \rm m}$ (top), and the fraction of extended emission at 13.2$\,\mu$m, FEE$_{13.2\,\mu m}$ (bottom), for individual galaxies in the GOALS sample. Galaxies are color-coded as a function of the 63/158\,$\mu$m ratio (top) and the strength of the silicate feature, \SSi\, (bottom). A linear fit to the datapoints without limits is shown in the top panel as a solid line. The dotted lines are the $\pm\,1\,\sigma$ uncertainty. We note that the 15$\,\mu$ luminosities are measured within the \textit{Spitzer}/IRS LL slit while the mid-IR sizes were obtained from the SL module at 13.2$\mu$m (\citealt{DS2010b}). For very extended sources, the MIPS 24$\mu$m images were used instead to measure the size of the starburst region. The intrinsic sizes (FWHM$_{\rm int}$) of the mid-IR emission were obtained after subtracting, in quadrature, the contribution of the instrumental profile (FWHM$_{\rm PSF}$) from the measured FWHM.}\label{f:ciifirvscompact}
\vspace{.5cm}
\end{figure}

It has been shown that the compactness of the mid-IR continuum emission of LIRGs (evaluated at $\lambda_{\rm rest}\,=\,13.2\,\mu$m) is related to their merger stage, mid-IR AGN-fraction and most importantly, to their far-IR color (\citealt{DS2010b}). LIRGs with higher \textit{IRAS} $S_\nu\,$60\,$\mu$m/$S_\nu\,$100\,$\mu$m ratios are increasingly compact. In other words, for a given \LIR, the dust in sources with far-IR colors peaking at shorter wavelengths is not only hotter but also confined towards a smaller volume in the center of galaxies. In \S\ref{ss:warm} we found that the \CIIno/FIR\, ratio is related to the average $T_{\rm dust}$ of our galaxies. Thus, we should expect to see a correlation between the \CIIno\, deficit and the luminosity surface density and compactness of LIRGs in the mid-IR. This is shown in Figure~\ref{f:ciifirvscompact}, where a clear trend is found for galaxies with higher luminosity surface densities at 15$\,\mu$m, $\Sigma_{15\,\mu \rm m}$ (top panel), or small FEE$_{13.2\,\mu m}$ (bottom panel), i.e, more compact, to show lower \CIIno/FIR\, ratios, irrespective of the origin of the nuclear power source.


Excluding those LIRGs for which only upper limits on their mid-IR size or \LCIIno\, are available,
we perform a linear fit to the data and obtain the following parameters for the correlation between \CIIno/FIR\, and $\Sigma_{15\,\mu \rm m}$:

\begin{equation}
log(\frac{\CIIno}{\rm FIR})\,=\,0.84(\pm\,0.27)\,-\,0.35(\pm 0.03)\,\times\,log(\Sigma_{\rm 15\,\mu \rm m})
\end{equation}\label{e:defvslsd}

\noindent
with a dispersion in the y-axis of 0.23\,dex.

\subsection{The Role of Active Galactic Nuclei}\label{s:agn}

It is known that the contribution of an AGN to the IR emission in LIRGs increases with \LIR\, (\citealt{Veilleux1995}; \citealt{Desai2007}; \citealt{Petric2011}; \citealt{AAH2012}). This is most noticeable at mid-IR wavelengths (\citealt{Laurent2000}; \citealt{Armus2007}; \citealt{Mullaney2011}) but a non-negligible fraction of the far-IR emission of ULIRGs can also be powered by an AGN. The EW of mid-IR PAH features is a simple diagnostic that has been widely used for the detection of AGN activity in galaxies at low and high redshifts (\citealt{Genzel1998}; \citealt{Armus2007}; \citealt{Desai2007}; \citealt{Spoon2007}; \citealt{Pope2008}; \citealt{Murphy2009}; \citealt{MD2009}; \citealt{Veilleux2009}; \citealt{Petric2011}; Stierwalt et al. 2013a). Broadly speaking, the PAH EW decreases as a component of hot dust at T$\,\gtrsim\,300\,$K, normally ascribed to an AGN, starts to increasingly dominate the mid-IR continuum emission of the galaxy. In addition, the hard radiation field of an AGN could be able to destroy a significant fraction of the smallest PAH molecules (\citealt{Voit1992}; \citealt{Siebenmorgen2004}). In particular, a galaxy is regarded as mid-IR AGN-dominated when its \PAHc\, EW $\lesssim\,0.3$, and it is classified as a pure starburst when \PAHc\, EW $\gtrsim\,0.5$ (although these limits are not strict). Sources with intermediate values are considered composite galaxies, in which both starburst and AGN may contribute significantly to the mid-IR emission.

\subsubsection{\CII\, Deficit in Pure Star-Forming LIRGs}\label{ss:definsb}

\begin{figure}[t!]
\vspace{.25cm}
\epsscale{1.15}
\plotone{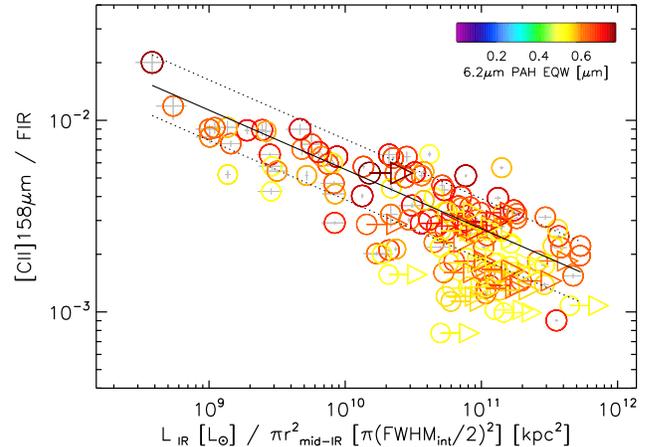}
\vspace{.25cm}
\caption{\footnotesize \CII/FIR\, ratio as a function of the nuclear \LIR\, divided by the area of the mid-IR emitting region ($\Sigma_{\rm IR}\,=\LIR\,/\pi r_{\rm mid-IR}^2$) for individual galaxies in the GOALS sample. This Figure is the same as Figure~\ref{f:ciifirvscompact} but using the nuclear \LIR\, of galaxies (scaled as the far-IR flux; see \S\ref{ss:dataanalysis}) instead of their 15$\mu$m monochromatic luminosity, and it is color coded as a function of the \PAHc\, EW. Only pure star-forming LIRGs, defined as to have \PAHc\,$\geq\,0.5\,\mu$m, are shown. The solid line is a fit to the data. See Eq.~(4).}\label{f:ciifirvslsdpah}
\vspace{.5cm}
\end{figure}

Figure~\ref{f:ciifirvslsdpah} shows the \CIIno/FIR\, ratio as a function of the IR luminosity surface density, $\Sigma_{\rm IR}$, of the LIRGs in GOALS, color-coded as a function of their \PAHc\, EW. As we can see, when only pure star-forming galaxies are considered (\PAHc\, EWs $\geq\,0.5\,\mu$m), the \CIIno/FIR\, ratio drops by an order of magnitude, from $10^{-2}$ to $\sim\,10^{-3}$. This indicates that the decrease in \CIIno/FIR\, among the majority of LIRGs is not caused by a rise of AGN activity but instead is a fundamental property of the starburst itself. It is only in the most extreme cases, when \CIIno/FIR\,$<\,10^{-3}$, that the AGN could play a significant role. In fact, powerful AGN do not always reduce the \CIIno/FIR\, ratio, as shown also in \cite{Sargsyan2012}. \cite{Stacey2010} also find that the AGN-powered sources in their high-redshift galaxy sample display small \CIIno/FIR\, ratios. However, they speculate that except for two blazars, the deficit seen in these sources could be caused compact, nuclear starbursts (with sizes less than $1-3\,$kpc) perhaps triggered by the AGN. We will return to this discussion in \S\ref{ss:definagn}.

The result obtained above also implies that the \CII\, line alone is not a good tracer of the SFR in most local LIRGs since it does not account for the increase of warm dust emission (Figure~\ref{f:ciifirvsf63f158}) seen in the most compact galaxies that is usually associated with the most recent starburst. In Figure~\ref{f:ciifirvslsdpah} we fit the data to provide a relation between the \CIIno/FIR\, ratio and the $\Sigma_{\rm IR}$ for pure star-forming LIRGs. The analytic expression of the fit is:

\begin{equation}
log(\frac{\CIIno}{\rm FIR})\,=\,0.84(\pm\,0.21)\,-\,0.32(\pm 0.03)\,\times\,log(\Sigma_{\rm IR})
\end{equation}

\noindent
with a dispersion in the y-axis of 0.16\,dex. The slope and intercept of this trend are indistinguishable (within the uncertainties) from those obtained in Eq.~(3), which was derived by fitting all data-points including low \PAHc\, EW sources with measured mid-IR sizes. This further supports the idea that the influence of AGN activity is negligible among IR-selected galaxies with $10^{-3}\,<\,$\CIIno/FIR\,$<\,10^{-2}$ and that the increase in IR luminosity of these sources is due to a boost of their warm dust emission.

\subsubsection{The Influence of AGN in the \CIIno\, Deficit}\label{ss:definagn}

\begin{figure}[t!]
\vspace{.25cm}
\epsscale{1.15}
\plotone{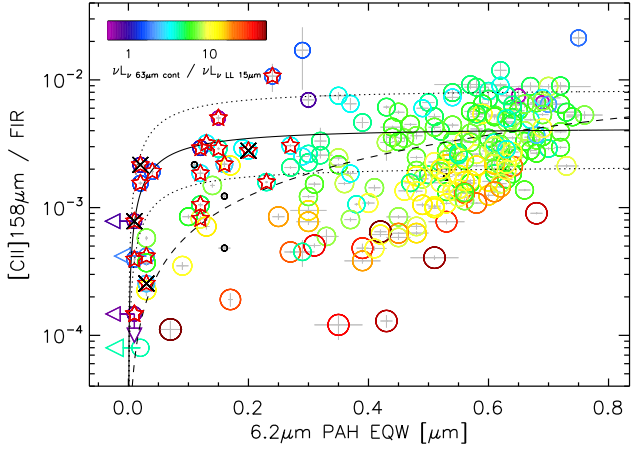}
\vspace{.25cm}
\caption{\footnotesize \CII/FIR\, ratio for individual galaxies in the GOALS sample as a function of the \PAHc\, EW measured with \textit{Spitzer}/IRS. Galaxies are color-coded as a function of their $\nu L_\nu\,$63/15\,$\mu$m ratio. If this information is not available, galaxies are shown as small black circles. The solid line represents the range in \CIIno/FIR\, and \PAHc\, with increasing contribution from an AGN (see text for details). The dotted lines are $\times\,2$ and $\times\,0.5$ the predicted trend, which accounts for variations in the \CIIno/\PAHc\, ratio and the \LFIR/$\nu L_{\nu\,{\rm 6\,\mu m}}$\, relation for star-forming galaxies. The dashed line assumes a decreasing of the \CIIno/\PAHc\, ratio proportional to the \PAHc\, EW due to pure PAH destruction from an AGN.}\label{f:ciifirvspahew}
\vspace{.5cm}
\end{figure}

Figure~\ref{f:ciifirvspahew} shows the \CII/FIR\, ratio as a function of the \PAHc\, EW for the LIRGs in GOALS. Starburst sources with large PAH EWs have a mean \CIIno/FIR\, ratio of $4.0\,\times\,10^{-3}$ with a standard deviation of $2.6\,\times\,10^{-3}$. As the \PAHc\, EW becomes smaller the dispersion increases and we find galaxies with both very small ratios as well as sources with normal values (or slightly lower than those) typical of purely star-forming sources (see also \citealt{Sargsyan2012}).

If AGNs contribute significantly to the far-IR emission of LIRGs and/or suppress PAH emission via photo-evaporation of its carriers, we would expect AGN-dominated sources to show significantly low \CIIno/FIR\, ratios and small PAH EWs. We have used the \textit{Spitzer}/IRS spectra of our galaxies to identify which of them are hosting an AGN based on several mid-IR diagnostics: (1) \NeV/\NeII\,$\,>\,0.5$; (2) \OIV/\NeII\,$\,>\,1$; (3) $S_\nu\,$30\,$\mu$m/$S_\nu\,$15\,$\mu$m $<\,6$ (i.e., $\alpha\,>\,-2.6$ for $S_\nu\,\propto\,\nu^\alpha$); as well as (4) the \PAHc\, EW itself (see constraints above). All these thresholds are rather restrictive and ensure that the contribution of an AGN to the mid-IR luminosity of a galaxy is \textit{at least} $25-50$\%. In Figure~\ref{f:ciifirvspahew} we mark those sources that have at least two positive indicators of AGN activity as red stars. We note that this excludes sources with low \PAHc\, EWs and no other AGN signatures, and is a more conservative cut than applied in \cite{Petric2011} to identify potential AGNs. Strikingly, these sources are not preferentially found at the bottom-left of the parameter space but instead as many as 2/3 show \CIIno/FIR\,$>\,10^{-3}$, typical of star-forming sources with large \PAHc\, EWs. This suggests that the impact of the AGN on the far-IR luminosity of these mid-IR dominated AGN LIRGs is very limited, unless the it contributes to both the \CIIno\, and far-IR in the same relative amount as the starburst does.

While $\sim\,18$\% of our sample appears to have significant AGN contribution to the mid-IR emission (\citealt{Petric2011}), the fraction in which the AGN dominates the bolometric luminosity of the galaxy is much smaller. 
To investigate this, we use two of the indicators described above, the \OIV\, line and the \PAHc\, EW, and the formulation given in \cite{Veilleux2009} to calculate the bolometric AGN fraction of those galaxies with at least two mid-IR AGN detections. We find that only four (20\%) of these galaxies have contributions $>\,50\,$\% in both indicators (black crosses in Figure~\ref{f:ciifirvspahew}).
Two galaxies have \CIIno/FIR\,$<\,10^{-3}$ (33\%) and two (14\%) a larger ratio.

To quantitatively asses the relationship between AGN activity and the \CIIno/FIR\, ratio among galaxies hosting an AGN, it is important to estimate first the AGN contribution to the far-IR flux. If we assume that the ratio of \CII\, to \PAHc\, emission of the star-forming LIRGs in GOALS is constant, as is the case for most normal, lower luminosity galaxies (\citealt{Helou2001}; \citealt{Croxall2012}; \citealt{Beirao2012}), we can calculate the expected evolution of the \CIIno/FIR\, ratio as a function of the \PAHc\, EW if we also assume that pure starbursts have a typical \PAHc\, EW$_{\rm SB}$\,=\,0.65$\,\mu$m and \CIIno/FIR\,=\,$4.0\,\times\,10^{-3}$ (as shown above), and that the average \LFIR/$\nu L_{\nu\,{\rm 6\,\mu m}}$\, ratios for pure star-forming galaxies and AGNs are $\sim\,$15 and $\sim\,$1, respectively. The value for star-forming galaxies varies from $\sim\,12-25$ in our sample while the value for AGNs has been estimated from the intrinsic AGN SED of \cite{Mullaney2011}. The predicted trend is shown in Figure~\ref{f:ciifirvspahew} as a solid black line, which agrees very well with the location of the AGNs identified with at least two mid-IR indicators (red stars). Under these assumptions, the \PAHc\, EW has to be reduced by a factor of $\sim\,$15 with respect to the \PAHc\, EW$_{\rm SB}$, i.e., down to $\simeq\,0.05\,\mu$m, before the AGN can contribute 50\,\% to the \LFIR. In fact, 2/3 of galaxies with EWs lower than this threshold have been identified as harboring an AGN by two or more mid-IR diagnostics. Therefore, 
only when the AGN contribution to the far-IR flux is significant do we see a noticeable decrease of the \CIIno/FIR\, ratio (always $<\,10^{-3}$).
We note however that the contrary might not be necessarily true since there are galaxies with low \CIIno/FIR\, ratios but with \PAHc\, EWs $\gtrsim\,0.5\,\mu$m.

We emphasize that this prediction does not account for possible destruction of PAH molecules due to the AGN. However, if the reduction of the PAH EW was entirely due to this effect, we would expect a linear correlation between the \CIIno/FIR\, ratio and the \PAHc\, EW, which is described by the dashed line in Figure~\ref{f:ciifirvspahew}. As we can see, the mid-IR AGN dominated galaxies do not follow the predicted trend, suggesting that PAH destruction is not important in LIRGs with \CIIno/FIR\,$\gtrsim\,10^{-3}$ at least at the scales probed by \textit{Herschel} and \textit{Spitzer}, in agreement with the results obtained in \cite{DS2011}.

Nearly half of galaxies with \CIIno/FIR\,$<\,10^{-3}$ and \PAHc\, EW $<\,0.05\,\mu$m have no other direct mid-IR diagnostic that reveals the presence of an AGN. Interestingly, all of them are among the outliers found in Figure~\ref{f:ciifirvssiabs}, showing an excess in the \SSi\, with respect to their observed \CIIno/FIR. We argued in \S\ref{ss:siabs} that these galaxies are probably hosting an extremely warm and compact source, optically and geometrically thick, not associated with the star-forming regions producing the bulk of the \CIIno\, and far-IR. The energy source of this component is unknown, though, since both an AGN or an ultra compact \HII\, region could generate such mid-IR signatures. However, the monochromatic $\nu L_\nu\,$63/15\,$\mu$m ratios displayed by these objects are $\gtrsim\,5$ (see color-coding in Figure~\ref{f:ciifirvspahew}), significantly higher than the typically flat spectrum seen in QSOs and pure AGN sources (\citealt{Elvis1994}; \citealt{Mullaney2011}) in which the hot dust emission dominates the mid-IR wavelengths up to $\sim\,30-50\,\mu$m, with $\nu L_\nu\,$=\,constant, and fading beyond. This adds evidence to the result obtained above that this type of deeply embedded objects only dominate the luminosity of the galaxy in the mid-IR.

Furthermore, we would like to emphasize that the fact that the source of this warm, compact emission does not produce the detected PAH or \CIIno\, emissions rules out models where PAH obscuration is invoked to explain the low PAH EWs found in these sources, since their observed \CIIno\, flux compared to that of the far-IR is also very low, implying that it is not extinction but rather the fact that the PDR emission of the warmest dust component in these LIRGs is actually extremely limited.

\section{Implications for Intermediate and High-Redshift Galaxy Surveys}\label{s:highz}

At intermediate redshifts, $z\,\sim\,1-3$, it has been found that IR-luminous galaxies span a wide range in \CIIno/FIR\, ratios: $\sim\,10^{-2}\,-\,10^{-3.5}$ (\citealt{Stacey2010}). A surprising discovery came from the most luminous systems, and the fact that many of them show values of this ratio similar to those found in local, lower luminosity galaxies (e.g., \citealt{Maiolino2009}; \citealt{HD2010}; \citealt{Sturm2010}; \citealt{Stacey2010}). These results, added to a number of recent findings obtained from the analysis of mid-IR dust features of star-forming galaxies using \textit{Spitzer}/IRS spectroscopy (e.g., \citealt{Pope2008}; \citealt{Murphy2009}; \citealt{Desai2009}; \citealt{MD2009}; \citealt{DS2010b, DS2011}; \citealt{Rujopakarn2011}; Stierwalt et al. 2013a) are pointing towards an emerging picture in which the local counterparts of the dominant population of IR-bright galaxies at intermediate and high redshifts ($z\,>\,1$) are not extremely dusty systems with similar IR luminosities (i.e., local ULIRGs) but rather galaxies with more modest SFRs, or \LIR\,$\simeq\,10^{10-12}$\,\Lsun\, (starbursts and LIRGs). 
Therefore, since GOALS is a complete, flux-limited sample of 60$\,\mu$m rest-frame selected LIRGs systems in the local Universe covering an IR luminosity range from $\sim\,10^{10}$ to $\sim\,10^{12}$\,\Lsun, the empirical relations we find can be used to estimate what might be seen in similar surveys of intermediate and high redshift IR-luminous galaxies.

Recently, \cite{Elbaz2011} has found that the majority of star-forming galaxies, from the nearby Universe and up to $z\,\sim\,2$, follow a "main sequence" (MS) that is depicted by a specific SFR (SSFR$_{\rm MS}$) that increases with redshift.
Galaxies in the MS are characterized by producing stars in a quiescent mode, with very low efficiencies (\LIR/\Mgas\,$\lesssim$\,10\,\Lsun/\Msun) and extended over spatial scales of several kpc (\citealt{Daddi2010a}; \citealt{Magdis2012}). On the other hand, galaxies with high SSFRs are currently experiencing a strong and very efficient, but short-lived (less than few hundred Myr) starburst event, some of them probably consequence of a major merge interaction.
The SSFR of local galaxies is anti-correlated with their compactness as measured in the mid-IR, as well as from radio wavelengths (see also Murphy et al. 2013). Therefore, the trends of \CIIno/FIR\, with FEE$_{13.2\,\mu m}$, $\Sigma_{15\,\mu \rm m}$, and $\Sigma_{\rm IR}$ found in \S\ref{ss:compact} and \S\ref{ss:definsb} tell us not only about the compactness of the starburst region but also about the main mode of star formation itself. Sources with high \CIIno/FIR\, ratios should belong to the MS while galaxies with low ratios will likely be compact, starbursting sources.

Because $>\,80-90$\% of the UV and optical light of star-forming LIRGs and ULIRGs is reprocessed by dust into the IR wavelengths (\citealt{Howell2010}), the \LIR/\Mstar\, ratio is equivalent to their SSFR since the SFR is directly proportional to the IR luminosity, with SFR$_{\rm IR}$/\LIR\,=\,1.72$\,\times\,10^{-10}\,$\Msun\,yr$^{-1}$\,\Lsun$^{-1}$ (as derived in \citealt{Kennicutt1998a}). Using Eq.~(13) from \cite{Elbaz2011} we calculate that the SSFR of MS galaxies in the local Universe is SSFR$_{\rm MS}\,\simeq\,0.09\,$Gyr$^{-1}$. However, this equation does not take into account the dependence of the SSFR$_{\rm MS}$ on the stellar mass of galaxies. Thus, we combine it with the SFR vs. \Mstar\, correlation obtained from the SDSS sample by \cite{Elbaz2007}, using a power-law index of 0.8 for the dependence of the SFR on \Mstar\, (their Eq.~(5)) and after normalizing it by the SSFR$_{\rm MS}$ at $z\,=\,0$.
This joint equation assumes that the exponential dependence of the SFR$_{\rm MS}$ as a function of \Mstar\, does not vary with $z$, which is roughly the case at least up to $z\,\sim\,2-3$ (see, e.g., \citealt{Karim2011}).
We have used this normalization factor to derive the excess of SSFR in our galaxies (also called "starburstiness": SSFR/SSFR$_{\rm MS}$).
If we define starbursting galaxies as those having a SSFR$\,>\,3\,\times\,$SSFR$_{\rm MS}$, then $\sim\,$68\,\% of the galaxies in GOALS would be classified as such.

\begin{figure}[t!]
\vspace{.25cm}
\epsscale{1.15}
\plotone{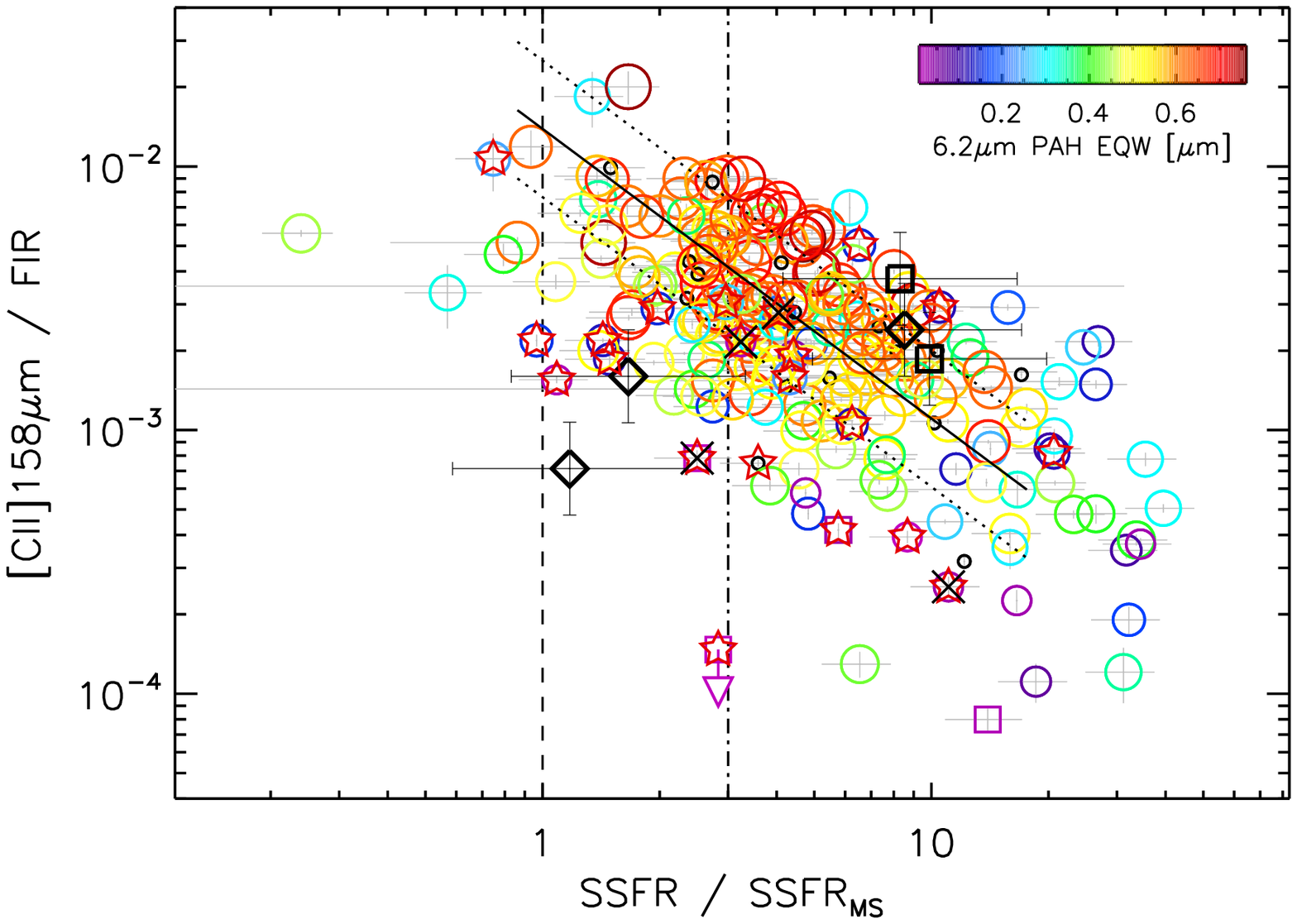}
\vspace{.25cm}
\caption{\footnotesize \CII/FIR\, ratio versus integrated SSFR normalized to the SSFR$_{\rm MS}$($z\,\sim\,0$) for individual galaxies in the GOALS sample. Galaxies are color-coded as a function of their \PAHc\, EW. Colored circles indicate sources for which an EW is available. Squares indicate lower limits. Small black circles are sources for which there is no information. The solid line is a fit to pure star-forming LIRGs only, excluding those sources that may harbor an AGN that could dominate their mid-IR emission (\PAHc\, EWs $<\,0.5\,\mu$m; see Figure~\ref{f:ciifirvspahew}). The dotted lines are the $\pm\,1\,\sigma$ uncertainty. The dashed line indicates the SSFR of MS galaxies at any redshift or \Mstar. The dotted-dashed line represents $3\,\times\,$SSFR$_{\rm MS}$, the limit above which galaxies are considered to be starbursting. The open squares represent two intermediate redshift galaxies at $z\,\sim\,1.2$, while the open diamonds represent three high redshift sub-millimeter galaxies at $z\,\sim\,4.5$ (see text for details).}\label{f:ciifirvsssfr}
\vspace{.5cm}
\end{figure}

Figure~\ref{f:ciifirvsssfr} shows the \CIIno/FIR\, ratio as a function of the integrated SSFR normalized to the representative SSFR$_{\rm MS}$ of galaxies at $z\,\sim\,0$ for our LIRG sample. The stellar mass values are taken from \cite{Howell2010} and were derived directly either from the 2MASS K-band or the 3.6$\,\mu$m IRAC luminosities using the $\Mstar/L$ conversions from \cite{Lacey2008}. The solid line represents a fit to pure star-forming galaxies with \PAHc\, EWs $\geq\,0.5\,\mu$m. The Pearson's test yields $r\,=\,-0.65$ ($p_r=0$), while the Kendall's test provides $\kappa\,=\,-0.47$. The correlation coefficient derived from the robust fit is $-0.76$.
We note that the SFR and \Mstar\, plotted here represent integrated measurements of our galaxies while the \CIIno\, and FIR values were obtained from a single PACS spaxel, probing an area of $\sim\,$9.4\arcsec\,$\times$\,9.4\arcsec, which at the median distance of our LIRGs is equal to a projected physical size of $\sim\,4\,$kpc on a side (similar to a 0.5\arcsec\, beam at $z\,\sim\,2$); an aperture big enough to contain most of the galaxies' emission. In fact, if a $3\,\times\,3$ spaxel aperture is used instead, the fit yield virtually the same parameters as those reported below (within the uncertainties).

The trend between \CIIno/FIR\, and normalized SSFR is not surprising since the IR luminosity appears in both quantities. Nevertheless, the correlation is indeed practical in terms of its predictive power. For example, we can see that there are no star-forming MS galaxies with \CIIno/FIR\,$<\,10^{-3}$. Their median ratio is $4.2\,\times\,10^{-3}$. On the other hand, starbursting sources show a larger range of ratios, from $10^{-2}$ to $10^{-4}$, with a median of $1.9\,\times\,10^{-3}$. The correlation between the \CIIno/FIR\, and the SSFR/SSFR$_{\rm MS}$ is given by the following equation:

\begin{equation}
log(\frac{\CIIno}{\rm FIR})\,=\,-1.86(\pm\,0.03)\,-\,1.09(\pm 0.07)\,\times\,log(\frac{SSFR}{SSFR_{MS}})
\end{equation}

\noindent
with a dispersion in the y-axis of 0.26\,dex. If the separation between MS and starbursting galaxies (high/low \CIIno/FIR) at any given redshift is related to an increase of the star formation efficiency (SFE; higher \LIR/\MHH; see, e.g., \citealt{Gracia-Carpio2011}; \citealt{Sargent2013}) then this equation can be applied to predict the \CIIno\, luminosity of star-forming galaxies at any $z$ for which a measurement of the (far-)IR luminosity and stellar mass are available as long as their SSFR is normalized to the SSFR$_{\rm MS}$ at that $z$ to account for the increase in gas mass fraction (\Mgas/\Mstar) and therefore SSFR of MS galaxies at higher redshifts (\citealt{Daddi2010a}; \citealt{Magdis2012}). Conversely, in future large IR surveys like those projected with the Cornell-Caltech Atacama Telescope (CCAT), this relation could be used for estimating the SSFR or stellar mass of detected galaxies when their \CIIno\, and far-IR luminosities are known.

As an example, in Figure~\ref{f:ciifirvsssfr} we show two intermediate redshift $z\,\sim\,1.2$ galaxies from \cite{Stacey2010} (SMMJ123633 and 3C368) and three high redshift $z\,\sim\,4.5$ sub-millimeter galaxies (SMGs) (LESS 61.1 and LESS 65.1: \citealt{Swinbank2012}; LESS 033229.3: \citealt{DeBreuck2011}; \citealt{Wardlow2011}) marked as open black squares and diamonds, respectively. To calculate the SFR of the SMGs we assumed that their \LIR\,$\simeq\,2\,\times$\,\LFIR. Since Eq.~(13) from \cite{Elbaz2011} is calibrated only up to $z\,\sim\,3$ and the dependence of the SSFR$_{\rm MS}$ on \Mstar\, is also uncertain beyond this redshift, we normalized the SSFR of the SMGs by the SSFR$_{\rm MS}$ at $z\,=\,3$. As we can see, three galaxies follow the correlation suggesting that they are starbursting sources with \CIIno/FIR\, ratios consistent with their normalized SSFRs.
On the other hand, the two remaining high-$z$ SMGs show significantly lower \CIIno/FIR\, ratios than the average of LIRGs for the same normalized SSFR. In particular, one of them display a \CIIno/FIR\, more than an order of magnitude lower than the value predicted by the fit to our local galaxy sample. Interestingly, both SMGs lie in the parameter space where most of the mid-IR identified AGN are located (red stars). This may suggest that these two galaxies could harbor AGN or unusually week \CIIno\, emission for their normalized SSFR.

Eq.~(3) and (4) can also be used to predict the mid- and total IR luminosity surface density of star-forming galaxies at high redshifts for which the \CIIno\, and far-IR fluxes are known, such as those that may be found in future spectroscopic surveys with the X-Spec instrument on CCAT.
With instantaneous coverage over all the atmospheric windows between 190 and 440~GHz, X-Spec will access the \CIIno\, line at redshifts from $\sim\,3.5$ to $\sim\,9$.
Moreover, if a measurement of the rest-frame mid-IR luminosity of galaxies is also available, this correlation can be further
used to estimate the physical size of the star-forming region in the mid-IR. This is particularly useful for z$\,\gtrsim\,3$ sources detected with \textit{Herschel} in deep fields. In this cases, the predicted mid-IR size of galaxies could be compared with direct measurements of the size of their far-IR emitting region as observed with ALMA on physical scales similar to those we are probing in our GOALS LIRGs with PACS.

Finally, because GOALS is a complete flux-limited sample of local LIRGs, we are able to predict the contamination of sources hosting AGNs in future large-scale surveys with both \CIIno\, and far-IR measurements. In Table~\ref{t:ciilfir} we provide the percentages of galaxies with mid-IR detected AGNs classified in different \CIIno/FIR\, and $S_\nu\,$63\,$\mu$m/$S_\nu\,$158\,$\mu$m bins. The values provided in the Table were computed using two conditions for the detection of the AGN that serve as upper and lower limits for the estimated fractions (columns (2) and (3)). The first was based on the \PAHc\, EW only and the second required of an additional mid-IR diagnostic to classify the galaxy as harboring an AGN (see above). For example, we predict an AGN contamination (based on the \PAHc\, EW only) of up to $\sim\,$70\% for \CIIno/FIR\,$<\,5\,\times\,10^{-4}$, which implies that at least $\sim\,$1/3 of IR-selected sources with extremely low \CIIno/FIR\, ratios will be powered by starbursts. Moreover, at the levels of \CIIno/FIR\,$\geq\,5\,\times\,10^{-4}$ or 63/158$\,\mu$m$\,<\,2$, the AGN detection fraction is expected to be $\lesssim\,20-25$\%.

\begin{deluxetable}{cccc}
\tabletypesize{\scriptsize}
\tablewidth{0pc}
\tablecaption{\scriptsize Fraction of AGN}
\tablehead{\colhead{\CIIno/FIR} & \colhead{AGN-frac} & \colhead{AGN-frac} & \colhead{63/158\,$\mu$m} \\ 
\colhead{range} & \colhead{\PAHc} & \colhead{multi} & \colhead{median} \\
\colhead{(1)} & \colhead{(2)} & \colhead{(3)} & \colhead{(4)}}
\startdata 
$>\,5\,\times 10^{-3}$ & 4\% & 2\% & 0.52 \\
$(1.5-5)\,\times 10^{-3}$ & 15\% & 10\% & 0.93 \\
$(0.5-1.5)\,\times 10^{-3}$ & 18\% & 9\% & 1.22 \\
$<\,5\,\times 10^{-4}$ & 72\% & 22\% & 1.92 \\
\hline
\hline
63/158\,$\mu$m & AGN-frac & AGN-frac & \CIIno/FIR \\
range & \PAHc & multi & median \\
(1) & (2) & (3) & (4) \\
\hline
$<\,0.5$ & 8\% & 4\% & 7.0$\,\times 10^{-3}$ \\
$0.5-1$ & 9\% & 8\% & 3.2$\,\times 10^{-3}$ \\
$1-2$ & 28\% & 14\% & 1.5$\,\times 10^{-3}$ \\
$>\,2$ & 56\% & 13\% & 5.8$\,\times 10^{-4}$
\enddata
\tablecomments{\scriptsize Top: (1) Range of \CIIno/FIR\, ratio; (2) Percentage of mid-IR detected AGN based only on the \PAHc\, EW of galaxies ($<\,0.3\,\mu$m; see \S\ref{s:agn}); (3) Percentage of mid-IR detected AGN based on multiple line and continuum emission diagnostics (see text); (4) Median $S_\nu\,$63\,$\mu$m/$S_\nu\,$158\,$\mu$m continuum flux density ratio within the range of \CIIno/FIR\, given in column (1).\\
Bottom: (1) Range of $S_\nu\,$63\,$\mu$m/$S_\nu\,$158\,$\mu$m continuum flux density ratio; (2) Percentage of mid-IR detected AGN based only on the \PAHc\, EW of galaxies (see text); (3) Percentage of mid-IR detected AGN based on multiple line and continuum emission diagnostics (see text); (4) Median \CIIno/FIR\, ratio within the range of $S_\nu\,$63\,$\mu$m/$S_\nu\,$158\,$\mu$m given in column (1).}\label{t:ciilfir}
\end{deluxetable}

\section{Conclusions}\label{s:summary}

We obtained new \textit{Herschel}/PACS \CII\, spectroscopy for 200 LIRG systems in GOALS, a 60$\,\mu$m flux-limited sample of all LIRGs detected in the nearby Universe. A total of 241 individual galaxies where observed in the \CII\, line. We combined this information together with \textit{Spitzer}/IRS spectroscopic data to provide the context in which the observed \CIIno\, luminosities and \CIIno/FIR\, ratios are best explained. We have found the following results:

\begin{itemize}

\item The LIRGs in GOALS span two orders of magnitude in \CIIno/FIR, from $\sim\,10^{-2}$ to $10^{-4}$, with a median of $2.8\,\times\,10^{-3}$. ULIRGs have a median of $6.3\,\times\,10^{-4}$. The \LCIIno\, range from $\sim\,10^7$ to 2$\,\times\,10^9\,\Lsun$ for the whole sample. The \CIIno/FIR\, ratio is correlated with the far-IR $S_\nu\,$63\,$\mu$m/$S_\nu\,$158\,$\mu$m continuum color. We find that all galaxies follow the same trend independently of their \LIR, suggesting that the main observable linked to the variation of the \CIIno/FIR\, ratio is the average dust temperature of galaxies, which is driven by an increase of the ionization parameter, $<$\textit{U}$>$.

\item There is a clear trend for LIRGs with deeper 9.7$\,\mu$m silicate strengths (\SSi), higher mid-IR luminosity surface densities ($\Sigma_{\rm MIR}$), smaller fractions of extended emission (FEE$_{13.2\,\mu m}$) and higher SSFRs to display lower \CIIno/FIR\, ratios. These correlations imply the the dust responsible for the mid-IR absorption must be directly linked to the process driving the observed \CIIno\, deficit. LIRGs with lower \CIIno/FIR\, ratios are more warm and compact (higher mid- and IR luminosity surface densities, $\Sigma_{\rm (M)IR}$), regardless of what is the origin of the nuclear power source. However, this trend is clearly seen also among pure star-forming LIRGs only, implying that it is the compactness of the starburst, and not AGN activity as identified in the mid-IR, that is the main driver for the declining of the \CIIno\, to far-IR dust emission. This implies that the \CIIno\, luminosity is \textit{not} a good indicator of the SFR in LIRGs with high \Tdust\, or large $\Sigma_{\rm IR}$ since it does not scale linearly with the warm dust emission most likely associated to the youngest stars.
There are a small number of LIRGs that have a larger \CIIno/FIR\, ratio than suggested by their deep \SSi\, and warm dust emission. The origin of the energy source of these LIRGs is unknown, although they likely contain a deeply buried, compact source with little or no PDR emission.

\item Pure star-forming LIRGs (\PAHc\, EW $\geq\,0.5$) have a mean \CIIno/FIR\,=\,$4.0\,\times\,10^{-3}$ with a standard deviation of $2.6\,\times\,10^{-3}$, while galaxies with low \PAHc\, EWs span the entire range in \CIIno/FIR. A significant fraction (70\,\%) of the LIRGs in which an AGN is detected in the mid-IR have \CIIno/FIR\, ratios $\geq\,10^{-3}$, similar to those of starburst galaxies suggesting that most AGNs do not contribute substantially to the far-IR emission. Thus, only in the most extreme cases when \CIIno/FIR\,$<\,10^{-3}$ might the AGN contribution be significant.
 
\item The completeness of the GOALS LIRG sample has allowed us to provide meaningful predictions about the \CIIno, $\Sigma_{\rm MIR}$, and AGN contamination of large samples of IR-luminous high-redshift galaxies soon to be observed by ALMA or CCAT. In a far-IR selected survey of high-$z$ LIRGs we expect to find up to $\sim\,$70\% of AGN contamination for \CIIno/FIR\,$<\,5\,\times\,10^{-4}$, which implies that at least 1/3 of IR-selected sources with extremely low \CIIno/FIR\, will be powered by starbursts. Moreover, above this ratio the AGN fraction is expected to be $\lesssim\,20-25$\%. For deep fields with \CIIno\, and far-IR emission measurements we can predict the IR luminosity surface density of galaxies, which could be compared with direct measurements of the size of their far-IR emitting region as observed with ALMA on physical scales similar to those we are probing in our GOALS LIRGs with PACS.


\end{itemize}

\section*{Acknowledgments}

We thank the referee for his/her useful comments and suggestions which significantly improved the quality of this paper. We also thank David Elbaz, Alexander Karim, J. D. Smith, Moshe Elitzur, and J. Gracia-Carpio for very fruitful discussions. L. A. acknowledges the hospitality of the Aspen Center for Physics, which is supported by the National Science Foundation Grant No. PHY-1066293. V. C. would like to acknowledge partial support from the EU FP7 Grant PIRSES-GA-2012-31578.  This work is based on observations made with the \textit{Herschel Space Observatory}, an European Space Agency Cornerstone Mission with science instruments provided by European-led Principal Investigator consortia and significant participation from NASA. The \textit{Spitzer Space Telescope} is operated by the Jet Propulsion Laboratory, California Institute of Technology, under NASA contract 1407. This research has made use of the NASA/IPAC Extragalactic Database (NED), which is operated by the Jet Propulsion Laboratory, California Institute of Technology, under contract with the National Aeronautics and Space Administration, and of NASA's Astrophysics Data System (ADS) abstract service.\\


\section{Appendix}\label{s:appendix}

\begin{figure}[t!]
\vspace{.25cm}
\epsscale{1.15}
\plotone{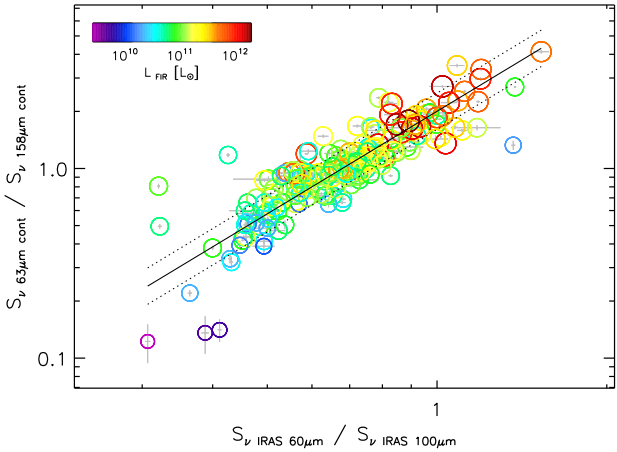}
\vspace{.25cm}
\caption{\footnotesize PACS-based $S_\nu\,$63\,$\mu$m/$S_\nu\,$158\,$\mu$m ratio as a function of \textit{IRAS}-based $S_\nu\,$60\,$\mu$m/$S_\nu\,$100\,$\mu$m ratio for individual galaxies in the GOALS sample. In LIRG systems with two or more galaxies, the same \textit{IRAS} color is plotted for each individual galaxy. The calculation of the \LFIR\, and the symbols are as in Figure~\ref{f:ciifirvsf63f158}.}\label{f:pacsvsiras}
\vspace{.5cm}
\end{figure}

In \S\ref{ss:warm}, Figure~\ref{f:ciifirvsf63f158}, we show the \CII/FIR\, ratio as a function of the PACS-based $S_\nu\,$63\,$\mu$m/$S_\nu\,$158\,$\mu$m ratio for the galaxies in the GOALS sample observed with \textit{Herschel}, and explain the reasons for adopting and plotting this far-IR color instead the more commonly used \textit{IRAS}-based $S_\nu\,$60\,$\mu$m/$S_\nu\,$100\,$\mu$m color. Here we show a comparison between both, to provide the reader with a tool for interpreting our results in terms of \textit{IRAS} colors, if necessary. Figure~\ref{f:pacsvsiras} shows that the far-IR ratios correlate well ($r\,=\,0.89$, $p_r\,=\,0$), as expected, with a slope of 1.80$\,\pm\,$0.08, an intercept of 0.30$\,\pm\,$0.02, and a dispersion in the y-axis of 0.10\,dex. We note that part of the scatter is caused by LIRG systems comprised by more than one galaxy, for which we have used their individual 63/158\,$\mu$m ratios but the same 60/100\,$\mu$m ratio, as they correspond to a single, unresolved \textit{IRAS} source. Nevertheless, independently of which of these far-IR colors we utilize, the same overall trend seen in Figure~\ref{f:ciifirvsf63f158} emerges, namely warmer galaxies display smaller \CIIno/FIR\, ratios. As mentioned in \S\ref{ss:warm}, when we use the \textit{IRAS} far-IR colors of integrated systems, our LIRG sample follows the same trend found for normal and moderate IR-luminous galaxies observed by \textit{ISO}. However, when using the 63/158\,$\mu$m ratio, the anti-correlation is tighter than that obtained when the emission from entire systems is employed, likely due to the fact that we are able to disentangle the true far-IR colors of individual galaxies. Moreover, the observed 63/158\,$\mu$m continuum ratios span a much larger dynamical range (a factor of $\sim\,$10, in contrast with the factor of $\sim\,3$ covered by the integrated 60/100\,$\mu$m ratios), which translates in a more accurate sampling of the average dust temperature of LIRGs.

\end{document}